\documentclass[galaxies,article,accept,moreauthors,pdftex]{Definitions/mdpi}

\firstpage{1}
\pubvolume{1}
\issuenum{1}
\articlenumber{0}
\pubyear{2021}
\copyrightyear{2020}
\datereceived{}
\dateaccepted{}
\datepublished{}
\hreflink{https://doi.org/} % If needed use \linebreak

\newcommand{\enzo}{\it{\small ENZO}}
\newcommand{\magcow}{\it{\small MAGCOW}}

\Title{MAGNETOGENESIS AND THE COSMIC WEB: A JOINT CHALLENGE FOR RADIO OBSERVATIONS AND NUMERICAL SIMULATIONS}
\TitleCitation{MAGCOW-The MAGnetised COsmic Web}

\Author{Franco Vazza $^{1,2,3}$, Nicola Locatelli$^{1,2}$, Kamlesh Rajpurohit$^{1,2}$, Serena Banfi$^{1,2}$, Paola Dom\'inguez-Fern\'andez$^{3,4}$,  Denis Wittor$^{3,1}$, Matteo Angelinelli$^{1,6}$, Giannandrea Inchingolo$^{1,2}$, Marisa Brienza$^{1,2}$, Stefan Hackstein$^6$, Daniele Dallacasa$^{1,2}$, Claudio Gheller$^{2}$, Marcus Br\"{u}ggen$^{3}$, Gianfranco Brunetti$^{2}$, Annalisa Bonafede$^{1,2}$, Stefano Ettori$^{5,6}$, Chiara Stuardi$^{1,2}$,Daniela Paoletti$^{5,6}$, Fabio Finelli$^{5,6}$}
\AuthorNames{Franco Vazza et al.  }
\AuthorCitation{Vazza F. et al.}
\address{%
$^{1}$ \quad Dipartimento di Fisica e Astronomia, Universit\'{a} di Bologna, Via Gobetti 93/2, 40129 Bologna, Italy\\
$^{2}$ \quad INAF-Istitituto di Radio Astronomia, via Gobetti 101, 40129 Bologna, Italy\\
$^{3}$ \quad Hamburger Sternwarte, Universit\"{a}t Hamburg, Gojenbergsweg 112, 41029 Hamburg, Germany\\
$^{4}$ \quad Department of Physics, School of Natural Sciences UNIST, Ulsan 44919, Korea\\
$^{5}$ \quad INAF, Osservatorio di Astrofisica e Scienza dello Spazio, via Gobetti 93/3, 40129 Bologna, Italy\\
$^{6}$ \quad INFN, Sezione di Bologna, viale Berti Pichat 6/2, I-40127 Bologna, Italy \\
$^7$ \quad Institute for Data Science, University of Applied Sciences of Northwestern Switzerland, Bahnhofstrasse 6, 5210 Windisch, Switzerland
}
\corres{Correspondence: franco.vazza2@unibo.it}

\abstract{The detection of the radio signal from filaments in the cosmic web is crucial to distinguish possible magnetogenesis scenarios. We review the status of the different attempts to detect the cosmic web at radio wavelengths. This is put into the context of the advanced simulations of cosmic magnetism carried out in the last few years by our  {\magcow} project.
While first attempts of imaging the cosmic web with the MWA and LOFAR have been encouraging and could discard some magnetogenesis models, the complexity behind such observations makes a definitive answer still uncertain.
A combination of total intensity and polarimetric data at low radio frequencies that the SKA and LOFAR2.0 will achieve is key to removing the existing uncertainties related to the contribution of many possible sources of signal along deep lines of sight. This will make it possible to isolate the contribution from filaments, and expose its deep physical connection with the origin of extragalactic magnetism.}


\begin{document}

\section{Introduction}
%...questions
\subsection{The puzzling origin of cosmic magnetism}
\label{subsec:intro1}

Understanding how the observed magnetic fields in the largest structures in the Universe emerged  is an important open issue in modern astrophysics.
%What is the origin of magnetic fields routinely detected  in the rarefied intergalactic space between galaxies by radio observations since the early 80's \cite[e.g.][]{2001ApJ...548..639K,fe12,bj14,2019SSRv..215...16V}?
%...dynamo
Theoretical and numerical work has shown that the microgauss ($\mu G$)  magnetic fields inferred from radio observations \cite[e.g.][]{2001ApJ...548..639K,fe12,bj14,2019SSRv..215...16V} could be produced by small-scale turbulent dynamo amplification, provided that the magnetic Reynolds number in the intracluster medium (ICM{\footnote{The full list of acronyms used in this work is given in the Abbreviations section at the end of the paper.}}) is large enough
 \cite[e.g.][]{2000ApJ...538..217C,2004ApJ...612..276S,ry08,2019JPlPh..85d2001R}. In typical ICM conditions (density $n \geq 10^{-4}$ [cm${}^{-3}$], temperature $T \sim 10^7-10^8$~K and plasma beta $\beta_{\rm pl} \sim 10^2$),  a dynamo can bring the magnetic energy density close to equipartition with the turbulent plasma kinetic energy in galaxy clusters. A small-scale dynamo also erases the memory of the initial magnetic field level, as verified by direct numerical simulations \cite[e.g.][]{do99,2012PhRvE..85b6303S,cho14,va18mhd,review_dynamo,dom19,2019ApJ...883..138R,2020MNRAS.494.2706Q,2021arXiv210807822S}.
 While all previous results are based on ideal magneto-hydrodynamical (MHD) simulations, kinetic simulations have recently started to explore the more realistic situation of weakly collisional plasmas. They confirmed the action of a dynamo but also highlighted the important role of Prandtl number variations on the initial magnetisation value and of kinetic instabilities \cite[e.g.][]{2014MNRAS.440.3226M,2016PNAS..113.3950R,2020JPlPh..86e9003S}.

 Regardless of the exact nature of the small-scale dynamo process in the ICM, a non-zero {\it seed magnetic field} is necessary for the dynamo to start. Then
 different sources and mechanisms to generate and amplify magnetic fields in diluted cosmic plasmas are viable.
 At present, we still do not know whether the observed fields have mostly a “primordial” origin (meaning they are the result of processes that happened soon after the Big Bang) or are  connected to the evolution of galaxies and of their black holes.

Several processes might have created truly {\it primordial} seed fields,  either during inflation by breaking the conformal invariance for the electromagnetic field, or by coupling it to other light fields and generating helicity \cite[e.g.][]{1988PhRvD..37.2743T,wi11,2018CQGra..35o4001H}. In the post-inflationary epoch,  a causal process could lead to the production of large amplitudes and small (below of the size of the Hubble radius at the generation time) coherence lengths \cite[][]{Vachaspati:1991nm,Caprini:2009yp,Ellis:2019tjf}.
 Recently, the first pilot cosmological simulations for the evolution of helical magnetic fields have been presented by \cite{2021arXiv210913520M}.
Assessing whether primordial magnetic fields were present at the epoch of the Cosmic Microwave Background (CMB) is important as they impact on the  Big Bang Nucleosynthesis, alter CMB anisotropies patterns and affect angular power spectra in temperature and polarisation \cite[e.g.][]{Paoletti:2008ck,Planck:2015zrl,Shaw:2009nf}.

%...astrophysical mechanisms
Alternatively or in addition, magnetic field seeds may have been injected by feedback events following the overcooling of gas onto the massive halos as well as the formation of stellar populations and supermassive black holes (SMBH). Such {\it astrophysical} sources can seed magnetic fields in cosmic structures at low redshift ($z \leq 10$), in an inside-out fashion which starts from galaxies: star formation drives winds of magnetised plasma into the circumgalactic medium \cite[e.g.][]{Kronberg..1999ApJ,Volk&Atoyan..ApJ.2000,donn09,2006MNRAS.370..319B,sam17} as well as into voids \cite[][]{beck13}. Jets and winds from active galactic nuclei (AGN) can magnetise the central regions of clusters and groups \cite[e.g.][]{2008A&A...482L..13D,xu09} and affect the transport of heat, entropy, metals and cosmic rays during the formation of cosmic structures \cite[e.g.][]{2016mssf.book...93P}.

Somewhat in between primordial and astrophysical scenarios,
additional processes such as the ``Biermann-battery'' mechanism \cite[][]{1997ApJ...480..481K}, aperiodic plasma fluctuations in the inter-galactic plasma \cite[][]{2012ApJ...758..102S},  resistive mechanisms \cite[][]{mb11}, ionization fronts around the first stars  \cite[][]{2005A&A...443..367L,2017MNRAS.472.1649D} or even the accretion of magnetic monopoles at the formation of primordial black holes \cite[][]{2021MNRAS.503.4387A} might provide additional seed magnetic fields.

%....putting it all together
Overall, it is fair to say that the present uncertainty in the amplitude of seed fields in the Universe, based on theory, is uncomfortably large. Basically, it allows for primordial fields with amplitudes in the range of $\sim ~10^{-34}-10^{-9}$ G \cite[e.g.][]{wi11,sub16,2021RPPh...84g4901V}. A recent survey of astrophysical seeding scenarios, with state-of-the-art cosmological MHD simulations, has inferred a lower limit of $B \sim 10^{-31} ~\rm G$ for the magnetisation of voids
\cite[][]{2021MNRAS.502.5726G}. Other limits have been suggested by modelling magnetic effects on CMB anisotropies from magnetically induced perturbations on post-recombination heating \cite[e.g.][]{Paoletti:2019pdi2}.
Such very low bounds seem to be incompatible with the absence of detected Inverse Compton Cascade emission from blazars \cite[e.g.][]{2009ApJ...703.1078D,2010Sci...328...73N,2015PhRvD..91l3514C,2015PhRvL.115u1103C,blazar21}, which yields limits of $\sim 10^{-16} \ \rm G$ {\footnote{See \cite{2012ApJ...752...22B} and \cite{2013ApJ...770...54M} for advanced discussions on the physical interpretations of such limits.}}. Other stringent limits have been suggested by modelling magnetic effects on post-recombination  heating \cite[e.g.][]{Kunze:2014eka,Chluba:2015lpa,Paoletti:2018uic} or by
the small-scale baryonic density fluctuation induced by primordial magnetic fields,
which would alter CMB anisotropies by promoting to inhomogeneous recombination and heating \cite[e.g.][]{2018MNRAS.481.3401T,2019PhRvL.123b1301J,2021arXiv210903816G}.
Recent limits for the average present-day magnetisation of the Universe have also been derived by the level of excess in diffuse radio emission detected by ARCADE2 and EDGES 21cm line experiments, yielding,$\leq 10^{-3}-0.3 \rm ~nG$ depending on the unknown spectral index of primordial seed fields \cite[][]{2021EPJC...81..394N}.

\begin{figure}[H]
\centering
\includegraphics[width=0.75\textwidth]{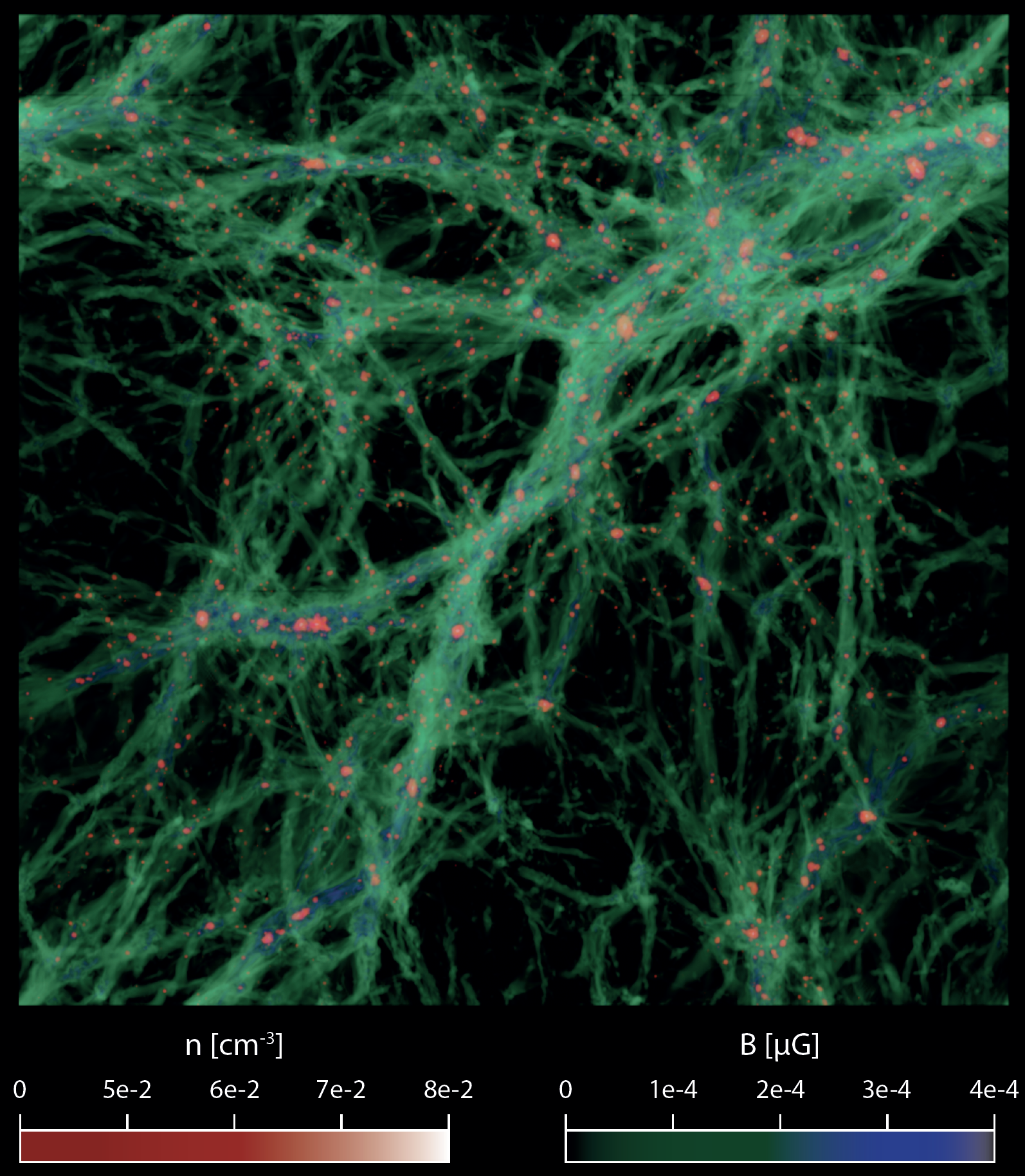}
\caption{3-dimensional volume rendering  of the gas density and of the magnetic field strength in one of the largest {\magcow} simulations of a primordial seed field ($B_0=0.1 \rm ~nG$ comoving) evolved in a $100^3 \rm Mpc^3$ volume using simulated with $2400^3$ cells and dark matter particles.}
\label{fig:chronos}
\end{figure}

\subsection{The magnetic cosmic web and its radio observations}
\label{radio_obs}

In the present situation, any significant detection of, or constraints on, magnetic fields beyond the scale of galaxies and galaxy clusters will have the potential of discriminating between different hypotheses for the origin of cosmic magnetism.
Previous work has shown that the radio signatures of drastically different magnetic field scenarios would leave very different imprints in Faraday Rotation and/or synchrotron emission from the magnetised cosmic web \cite[][]{donn09,va15radio,va17cqg}, in regions where small-scale dynamo is not expected to operate, due to the different local plasma conditions there \cite[e.g.][]{va14mhd,2018Galax...6..128L}.

In the (numerically) simple scenario in which magnetic fields were already in place at very early cosmic epochs, and got later advected, compressed and amplified up to the observed values within galaxy clusters and groups, we expect that the 3-dimensional distribution of the magnetic cosmic web pretty much follows that of the matter cosmic web.

Figure \ref{fig:chronos} shows a volume rendering \footnote{The volume rendering was realised using the CG Animation Software Blender in which we imported the ENZO simulation data previously converted in the openVDB format. This data format is optimised to maximise the shading of the volume rendering (and so the resolution and contrast of quantities' small details) using light tracing not reachable with standard scientific visualisation tools such as Paraview.} of baryon gas density and magnetic fields in one of the largest {\magcow} simulations (and cosmological MHD simulations in general), i. e. a comoving $100^3\,\rm Mpc^3$ volume simulated with $2400^3$ cells and dark matter particles, for a constant spatial comoving resolution of $41.6 \rm ~kpc$ and assuming a simple uniform seed magnetic field of $B_0=0.1 \rm ~nG$ comoving, seeded  at the beginning of the run ($z=40$).
For further analysis of this simulation see \cite{va19,vf20,vern21,figaro1}.
Spectacular threads of magnetic fields that trace the underlying matter distribution of the cosmic web are clearly seen in the image. These fields are the result of an initially uniform magnetic field distribution that has been compressed by the matter field. Given that the isotropic adiabatic compression of magnetic field lines increases the strength by $|B| \propto (\rho/\langle \rho \rangle)^{2/3}$ (where $\rho$ is the gas density and $\langle \rho \rangle$ is the gas mean density) in the absence of dynamo amplification the range of amplitude in magnetic field strength is smaller than the range of gas densities across the cosmic web. We anticipate that in astrophysical scenarios, on the other hand,  the distribution of magnetic fields is much more confined closer to the high density peaks of the cosmic web \cite[e.g.][]{va17cqg}.

The diffuse gas in the cosmic web, especially in the low redshift Universe ($z \leq 1$), has not been imaged in X-ray, owing to the scarcity of detectable X-ray photons from the warm-hot intergalactic medium \cite[e.g.][]{2021ExA...tmp...41S,2021arXiv210201096A}. The exceptions are attempts involving massive galaxy clusters \cite{2015Natur.528..105E,2021A&A...647A...2R}, or stacking techniques \cite[][]{2020A&A...643L...2T}. In the radio band, the situation may be less challenging because cosmic filaments are surrounded by strong accretion shocks \cite[e.g.][]{ry03,pf06}, which may also accelerate cosmic rays and produce synchrotron emission \cite[e.g.][]{2004ApJ...617..281K,2011JApA...32..577B,2012MNRAS.423.2325A,va15radio}.

In the last few years, new low-frequency radio telescope have detected magnetic fields beyond the scale of galaxy clusters.  In several cases our {\magcow} simulations were used to quantify the implications of such limits, and turn them into constraints on the different models of cosmic magnetism.

First, a few short ($\leq 3 \rm ~Mpc$)  and overdense ($\rho \sim 50-10^2 \langle \rho \rangle$) filaments between pre-merging clusters were detected by LOFAR-High Band Antenna (HBA): in Abell\,399-Abell\,401 \cite[][]{2019Sci...364..981G} and Abell\,1758A and B \cite[][]{2020MNRAS.499L..11B}. The modelling of the steep spectrum ($\alpha \leq -1.3$, where $I(\nu) \propto \nu^{\alpha}$) emission in Abell\,399-Abell\,401, which currently prefers a scenario based on turbulent re-acceleration of fossil relativistic electrons, suggested volume-filling $\sim 0.5~\mu G$ magnetic fields \cite{bv20}. A similar case, albeit more internal to the cluster volume, appears to be the bridge connecting the halo and relic region in the Coma cluster, whose recent LOFAR-HBA observation has also been modelled with our simulations \cite{bo21}.

On the other hand, LOFAR-HBA observations of two clusters connected by longer filaments ($\sim 10-20 \rm ~Mpc$, which are expected to enclose a $\rho \sim 5-50 \langle \rho \rangle$ gas overdensity) could not detect diffuse radio emission, suggesting a $\leq 0.25 ~\mu G$ upper limit for the magnetisation of such extended filaments, under the hypothesis that relativistic electrons are accelerated by structure formation shocks here \cite{lo21L}, .

On even larger scales, upper limits on the amplitude of magnetic fields have been statistically derived from the lack of detection of the cross-correlated signal between the large-scale distribution of galaxies in optical and IR surveys, and MWA surveys, which was used to suggest a $\leq 0.1~\mu G$ upper limit on the average magnetisation of the $z \leq 0.1$ cosmic web surrounding halos, i. e. $\rho \sim 10-10^2 \langle \rho \rangle$ \cite{brown17,vern17}.

Finally, \cite{vern21} have made a statistical detection of the diffuse radio emission between pairs of halos separated by $\leq 15 ~\rm Mpc$, with the stacking of  $390,808$ pairs of luminous red galaxies identified in the Sloan Digital Sky Survey Data Release 5 and observed at four different radio frequencies (from 50 to 120\,MHz) using the the GLEAM survey with MWA and with the Owens Valley Radio Observatory Long Wavelength Array. The reported detection is at the
$\sim 5 \sigma$ level, has a spectral index of $\alpha \approx -1.0$ and appears difficult to explain with any known population of radio galaxies and AGN. Based on the  analysis of our {\magcow} simulations, the emission was suggested to be compatible with a diffuse magnetic field in filaments, in the $\sim 10-20 \rm ~nG$ ballpark for a typical expected gas density in the $\rho \sim 5-50 \langle \rho \rangle$ range.

However, building a consistent mock statistical analysis with simulations is far from trivial, and  more recent work using the same {\magcow} simulations has suggested that it is difficult to reproduce the same excess signal from cluster pairs, if all outliers are removed \cite[][]{figaro2}.

A way to overcome sensitivity limitations of total intensity in radio observations is to use the Faraday Rotation Measure (RM) along the line-of-sight to a polarised background radio source. This can be used to perform a tomography of the cosmic web, with a fine grid of background sources \cite[e.g.][]{2014ApJ...790..123A, 2016A&A...591A..13V}. Fully exploiting this method currently suffers from the lack of large and dense grids of polarised sources, which should become available with the SKA \cite{2018Galax...6..128L}. Early attempts to constrain extragalactic magnetic fields statistically have set very high upper limits at the level of $0.3-7 \,\mu$G \cite[e.g.][]{Blasi.Burles..1999,2006ApJ...637...19X,2016A&A...591A..13V}.
Later work by \cite[][]{2016PhRvL.116s1302P} used the (lack of) evolution of RM in sources observed with the  NRAO VLA Sky Survey (NVSS) to infer much lower, $0.65-1.7 \rm ~nG$ limits for the extragalactic magnetic fields.
More recently, \cite{2019A&A...622A..16O} claimed the detection of the RM contribution by filaments overlapping the polarised emission by a giant radio galaxy using LOFAR-HBA.

By cross-correlating the RM signal of 1742 radio galaxies in the NVSS \cite[][]{2009ApJ...702.1230T} with the distribution of optical and IR $z \leq 0.5$ galaxies,  \cite{2021MNRAS.503.2913A} recently measured a $\leq 3.8 \rm ~rad/m^2$ upper bound to the RM contribution of the magnetised cosmic web, which may suggest a $\leq 30 \rm ~nG$ upper limit on the magnetisation of the cosmic web on $1-2.5 \rm ~Mpc$ scales ($\rho \sim 5-50 \langle \rho \rangle$).

The excess RM between random pairs of extended radio galaxies can be used to constrain the amplitude of intergalactic magnetic fields. This is done by comparing the $\Delta RM$ difference, as a function of angular scale, between physically related pairs of extended radio galaxies, and random associations of radio lobes \cite[][]{vern19,st20}.  In particular, \cite{os20} found that the RM difference between random and physical pairs of $349$ radio galaxies observed with LOFAR-HBA was $|\Delta RM^2| \leq 1.9 \rm ~rad^2/m^4$. Such a difference constrains the magnetisation of cosmic matter to $\leq 4 \rm ~nG$ for field with a coherence length $\leq \rm ~Mpc$,  and  $\rho \sim 1-10 \langle \rho \rangle$.

Finally, more indirect constraints on the amplitude of magnetic fields on the largest scales have been inferred from the Cosmic Microwave Background (see Sec. \ref{subsec:intro1}). These can be turned into an estimate on the current magnetisation of voids, under the reasonable hypothesis of ideal MHD and lack of other sources of magnetisation in the very low density Universe ($\rho \sim 10^{-2}-1 \langle \rho \rangle$). See Table \ref{table:tab1} for a summary of observational limits on cosmic magnetic fields.

 In Section \ref{results} we will attempt to translate such measurements into constraints on extragalactic magnetic fields on different scales, with the strong caveat that each different measurement comes from very different selection functions, class of sources and redshift selection. They are also very differently affected by cosmic foregrounds, background and observational limitations.
 In Section \ref{sec:conclusions} we present our conclusions and put them in the broader context of the aim to find the origin of cosmic magnetism.

\begin{table}[H]
\begin{center}
\caption{(Incomplete) list of recent and relevant measures or upper limits that attempted to detect magnetic fields on extragalactic scales, together with the estimated overdensity range their refer to.}
\footnotesize
\centering \tabcolsep 2pt
\begin{tabular}{c|c|c|c|c}
  Observation  & estimate on $|B|$ & approx. density range & instrument(s)  & references\\ \hline
Sync.in cluster bridges & $\sim 0.2-0.5 ~\mu G $& $\rho/\langle \rho \rangle \sim 50-200$ & LOFAR-HBA (120 MHz) & \cite{2019Sci...364..981G,bv20} \\

Sync. in cluster pairs & $\leq 0.25 \mu G$ & $ \rho/\langle \rho \rangle \sim 5-50 $ & LOFAR-HBA (120 MHz) & \cite{lo21L}\\

Optical-radio cross-corr. & $\leq 0.25 ~\mu G $ &$ \rho/\langle \rho \rangle \sim 10-10^2 $ & MWA-EoR0 (180 MHz) & \cite{vern17,brown17}\\

Sync. stacking of cluster pairs & $\sim 10-20 \rm nG $ &  $ \rho/\langle \rho \rangle \sim 5-50 $ & MWA+LWA (50-120 MHz) & \cite{vern21}\\

$\Delta RM (\theta)$ of radio gal. pairs & $\leq 40 \rm ~nG $&  $ \rho/\langle \rho \rangle \sim 1-10 $  & VLA-NVSS (1400 MHz) & \cite{vern19} \\

 $\Delta RM(\theta)$ of radio gal. pairs &$\leq 4 \rm ~nG$ &  $ \rho/\langle \rho \rangle \sim 1-10 $ & LOFAR-HBA (120 MHz) & \cite{os20,st20} \\

   $RM$ cross-correlation & $\leq 30 \rm ~nG$ & $ \rho/\langle \rho \rangle \sim 1-10 $  & VLA-NVSS (1400 MHz) & \cite{2021MNRAS.503.2913A} \\
    Excess $RM$ across $z$ & $\leq 1.7 \rm ~nG$ & $ \rho/\langle \rho \rangle \sim 1 $  & VLA-NVSS (1400 MHz) & \cite{2016PhRvL.116s1302P} \\
  CMB anisotropies T\&P & $\leq 2.8 \rm ~nG$ & $ \rho/\langle \rho \rangle \sim 1 $ & PLANCK2018+BK15+SPTPol &  \cite{Paoletti:2019pdi2}\\

 CMB heating & $\leq 0.83 \rm ~nG$ & $ \rho/\langle \rho \rangle \sim 1 $ & PLANCK-2015 & \cite{Paoletti:2018uic} \\

  Excess Sync. Radiation  & $\leq 10^{-3}-0.3 \rm ~nG$ & $ \rho/\langle \rho \rangle \sim 1 $  & ARCADE2+LW1 ($78 \rm ~MHz$)& \cite{2021EPJC...81..394N} \\
  Blazar Inv. Compton  & $\geq 10^{-7}-10^{-5} \rm nG$ &$ \rho/\langle \rho \rangle \sim 10^{-2}-1$ & VERITAS, HAWC, FERMI & \cite{2009ApJ...703.1078D,2010Sci...328...73N,2015PhRvD..91l3514C,2015PhRvL.115u1103C,blazar21}   \\

  \end{tabular}
  \end{center}
\label{table:tab2}
\end{table}

\section{Methods \& Materials: Cosmological Simulations of the Cosmic Web}
\label{methods}

The {\magcow} project, funded by a Starting Grant from the European Research Council \footnote{https://cosmosimfrazza.myfreesites.net/erc-magcow} is built around the idea of exploring the origin of cosmic magnetism, through the combination of radio observations capable of probing the most rarefied cosmic regions of the Universe, and advanced numerical simulations.

Over the latest few years we produced large simulations for the  co-evolution of dark matter (DM), gas matter, supermassive black holes and stellar populations, together with their associated magnetic fields, in an expanding space-time, with customized versions of the {\enzo} code \cite[][]{enzo14}.

This paper is not meant to be too detailed on the numerical subtleties of our simulations, which have been documented elsewhere \cite[][]{2019MNRAS.486..981G,Banfi20,gv20,va21}. Here, we review the most relevant physical aspects of the models examined in our old and new runs, and refer to the Appendix  for a more complete overview of the numerical details and code implementation specifics.

A first large set of (25) simulations has been already described in detail in \cite[][]{va17cqg}. In this contribution, we also include the results from more recent re-simulations, tailored to improve the comparison with new observational results.

 All models use some prescription for the seeding of magnetic fields, either at the start of the simulation, mimicking primordial mechanisms, or during the simulation, mimicking astrophysical generation of magnetic fields.

The older simulations included in this paper are (for consistency we keep the same naming given in previous papers):

\begin{itemize}
    \item P: a baseline primordial model in which we initialised a spatially uniform $B_0=1 \rm ~nG$ seed field at the start of the simulation;
     \item DYN5: a run starting from a negligible and spatially uniform $B_0=10^{-9} \rm ~nG$ seed field, in which we used a sub-grid model for turbulent dynamo amplification (see Appendix), for cells with a gas density $\rho \geq 2~\langle \rho \rangle$, i. e. basically already within filaments;
     \item CSF2: a run including radiative cooling, star formation and feedback, starting from a negligible and spatially uniform $B_0=10^{-9} \rm ~nG$ seed field. Magnetic fields are released at every episode of thermal feedback from star formation  (with a fixed small conversion efficiency per event, $\leq 10\%$), to reproduce an astrophysical seeding scenario in which winds from supernova remnant magnetise the large-scale structures.
     \item CSFBH2: similar to CSF2, but also including the formation, growth, merger and feedback events of supermassive black holes. Also in this case magnetic energy is released by SMBH feedback events ($\sim 10\%$ of the feedback energy) mimicking an astrophysical seeding scenario in which active galactic nuclei magnetise large-scale structures. The model parameters for baryon physics were tuned in previous work to reproduce cosmic star formation history and galaxy groups/clusters scaling relations  (see Appendix).
\end{itemize}

The new runs are introduced here for the first time:

\begin{itemize}

\item  P01: a simple primordial model in which we initialised a spatially uniform $B_0=0.1 \rm ~nG$ seed field at the start of the simulation, and in which we allowed for the sub-grid dynamo amplification of magnetic fields only for cells with a gas density $\rho \geq 50~\langle \rho \rangle$, which approximately marks region within the virial radius of halos, where  turbulence is predicted to be well developed and mostly solenoidal \cite[][]{miniati14}. As a further relevant difference with the ``old" DYN5, the amplification efficiency is increased by a factor 10 for $z \leq 1$, leading to more realistic magnetic fields in low-redshift clusters and groups.
\item CMB2, CMB3 \& CMB4: we simulated primordial magnetic fields derived from the constraints by the Cosmic Microwave Background observations following our recent work \cite[][]{va21}. In detail, the fields scale dependence is described by a power law spectrum: $P_B(k) = P_{B0}k^{\alpha} $ characterised by a constant spectral index and an amplitude, commonly referred  by smoothing the fields within a scale $\lambda=1 \rm ~ Mpc$, and therefore in the remainder of the paper we will use
$B_{\rm Mpc}$ to refer to the smoothed magnetic field amplitude. In this work, we used the three  models outlined in \cite{va21}, which resulted as the least challenged by low-redshift radio observations of the cosmic web: we assumed $\alpha=-1.0$ (CMB2), $0.0$ (CMB3) and $1.0$ (CMB4) and $B_{\rm Mpc}=1.87 $, $0.35$ and $0.042 \rm ~nG$ (comoving), respectively. All runs also adopted the same run-time sub-grid model for dynamo amplification, as in the P01 model.
\end{itemize}

 Table 2 summarises all relevant parameters of our runs and more information is given in the Appendix.

All runs assumed a $\Lambda$CDM cosmological model, with density parameters $\Omega_{\rm BM} = 0.0478$, $\Omega_{\rm DM} =
0.2602$,  $\Omega_{\Lambda} = 0.692$, and a Hubble constant $H_0 = 67.8$ km/sec/Mpc \cite[][]{2016A&A...594A..13P}.  Runs were started at $z=40$ and  have the constant spatial  resolution of  $83.3 ~\rm kpc/cell$ (comoving) and the constant mass resolution for DM of  $m_{\rm dm}=6.19 \cdot 10^{7}M_{\odot}$ per particle.

 Our simulations are limited to the ideal MHD scenario, meaning that the resistive dissipation of magnetic fields, kinetic plasma effects and other features connected with the departure from the single fluid model cannot be accounted for. Although it is fair to say that these phenomena have been scarcely studied in the very low-density regime which we are mostly concerned with here. The  numerical work that investigated non-ideal MHD effects reported negligible differences compared to ideal MHD in the periphery of galaxy clusters, or in even more rarefied environments \cite[e.g.][]{2011MNRAS.418.2234B,ruszkowski11}.

\begin{table}[H]
%\begin{center}
\caption{Main parameters of {\magcow} runs employed in this work. All quantities given are comoving. For more details, see Sec. \ref{methods} and the Appendix. }
\footnotesize
\centering \tabcolsep 2pt
\begin{tabular}{c|c|c|c|c|c|c|c}
  $N_{\rm grid}$ & $\Delta x$ & $L_{\rm box}$  & astroph. seeding & Initial $B_{\rm Mpc}$ & $\alpha_B$ & sub-grid dynamo  & ID\\
   &   [kpc] & [Mpc] & & [nG]  & & & \\  \hline
%  $2400^3$ & 41.6 & 100  & no & 0.1 & - & no & HR\\ \hline
        $1024^3$ & 83.3 & 85 & no  &  $1.0$ & -  & no  &  P\\
       $1024^3$ & 83.3 & 85 & no &  $10^{-9}$  & - & dynamo mod.1  & DYN5\\
       $1024^3$ & 83.3 & 85 & stellar feedback & $10^{-9}$ & - & no & CSF2\\
       $1024^3$ & 83.3 & 85 & stellar \& SMBH feedback & $10^{-9}$ & - & no & CSFBH2\\ \hline
       $1024^3$ & 83.3 & 85 & no  &  $0.1$ & -  & dynamo mod.2 &  P01\\
       $1024^3$ & 83.3 & 85 & no  & $1.87$ & $-1.0$   & dynamo mod.2  &  CMB2\\
       $1024^3$ & 83.3 & 85 & no  & $0.35$ & $0.0$   & dynamo mod.2  &  CMB3\\
       $1024^3$ & 83.3 & 85 & no & $0.042$ & $1.0$  & dynamo mod.2  &  CMB4\\
         \end{tabular}
\label{table:tab1}
\end{table}

\begin{figure}[H]
\centering
\includegraphics[width=0.36\textwidth]{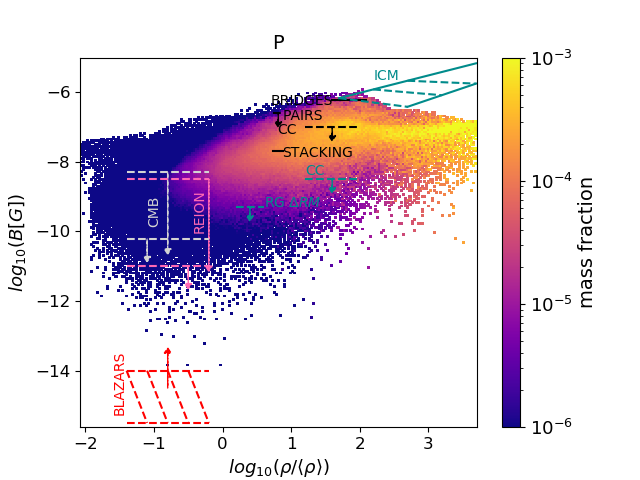}
\includegraphics[width=0.36\textwidth]{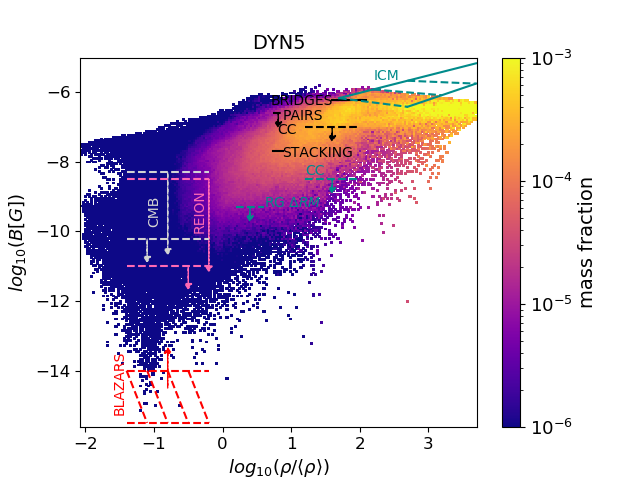}
\includegraphics[width=0.36\textwidth]{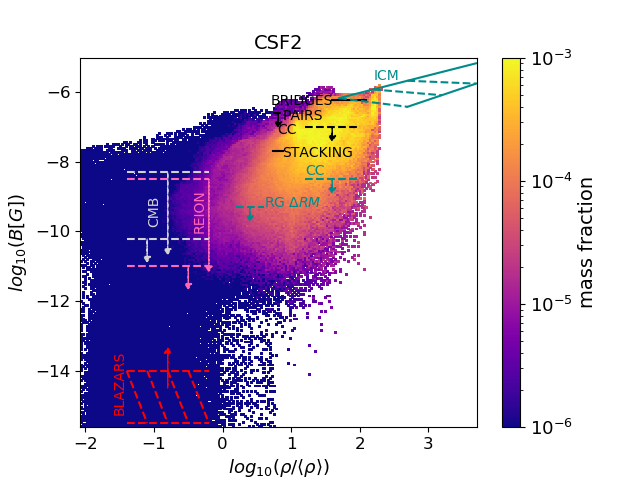}
\includegraphics[width=0.36\textwidth]{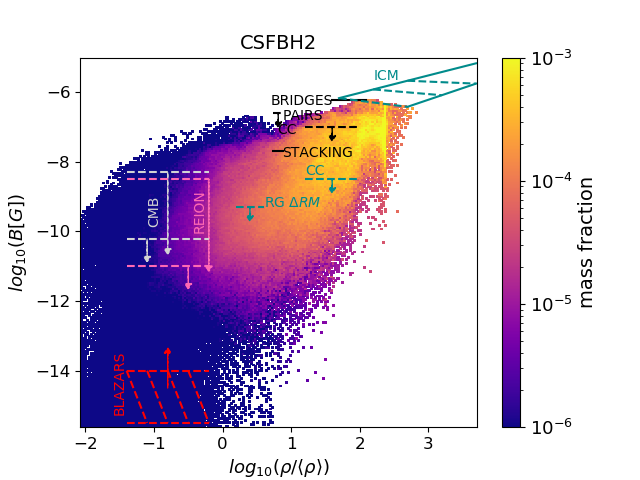}
\includegraphics[width=0.36\textwidth]{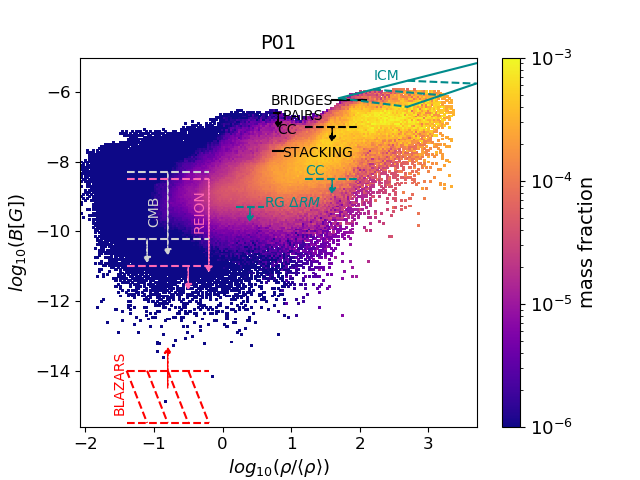}
\includegraphics[width=0.36\textwidth]{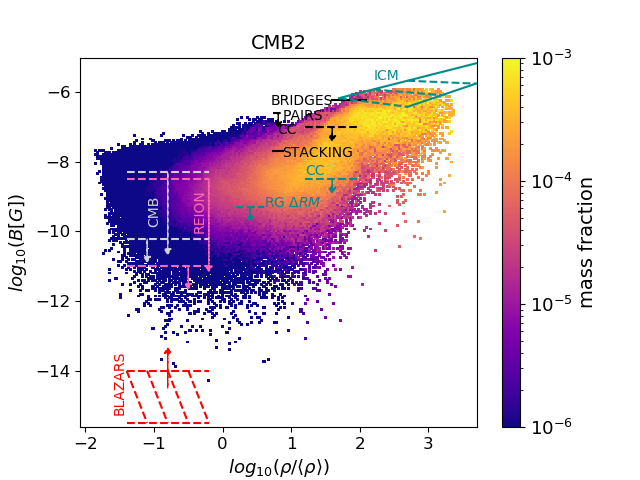}
\includegraphics[width=0.36\textwidth]{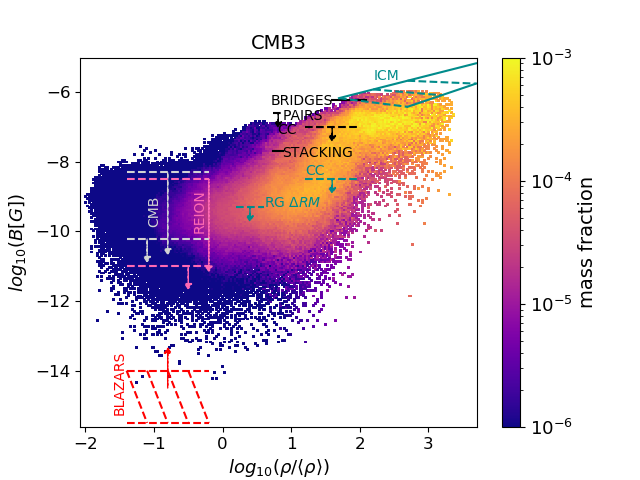}
\includegraphics[width=0.36\textwidth]{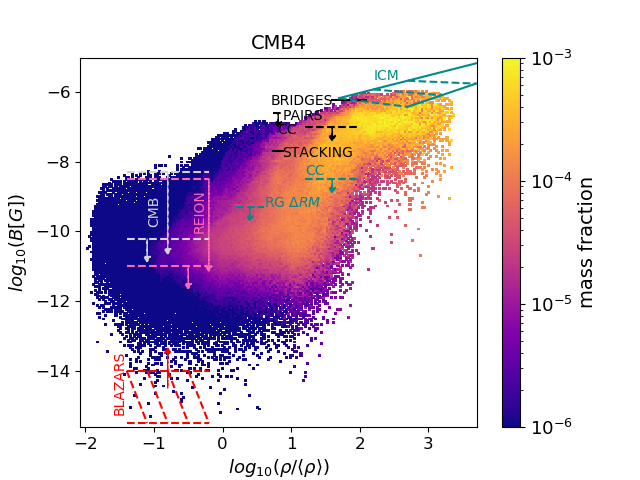}
\caption{Phase diagrams $(|B|, \rho/\langle \rho \rangle)$ for all {\magcow} runs discussed in this paper, at $z=0.02$. The contribution to each pixel is weighted by the gas mass fraction within a $85^3$ $\rm Mpc^3$ volume. We additionally overlay the approximate range of detections and limits by current observations, as in Tab.\ref{table:tab1}. The dark cyan colors represent Faraday Rotation observations, the black lines synchrotron total intensity observations, the light grey lines are measurements of CMB multipoles, the pink lines are for estimates based on reionisation, and the red lines are from blazars observations. Solid lines refer to detections and dashed lines refer to non-detections.}
\label{fig:phases}
\end{figure}

\section{Results}
\label{results}

We start by looking at the statistical distribution of magnetic field strength as a function of gas overdensity, $(|B|, \rho/\langle \rho \rangle)$. This was measured at $z=0.02$ in the old {\magcow} runs presented in Sec. \ref{methods}, which are the first four panels in Fig. \ref{fig:phases}. Each pixel in the phase diagram is weighted by the relative gas mass fraction.  We overlay onto the various phase distributions the approximate constraints that can be derived from observations.  According to our best estimate of the gas overdensity along the line-of-sight which is mostly probed by each observation, as in Sec. \ref{radio_obs}.

As expected, most cosmic environments carry the memory of magnetic seeding and show very different trends.
For example, in sheets and low-density filaments, $\rho \sim \langle \rho \rangle$, most of gas matter in the P model carries memory of the seed field, having $B \sim 1 \rm ~nG$. This is significantly lower in the DYN5 model owing to the partial effect of dynamo amplification here, which is assumed to operate even in very low density regimes. In the astrophysical scenarios (CSF2 and CSFBH2), the average magnetisation is much lower ($\leq 0.01 \rm ~nG$) and there is a $\sim 6$ orders of magnitude spread in the possible values of $|B|$. The latter depends on the proximity of cells to a nearby source of magnetisation.
Such differences persist in the denser environment of cluster outskirts, where the magnetic field strengths are better approximated by the adiabatic $|B| \propto (\rho/\langle \rho \rangle)^{2/3}$ relation.

On densities larger than the typical density enclosed by virialised  halos, all models struggle to match the magnetisation observed in real clusters and groups (which we show using the latest compilation of RM results in galaxy clusters by \cite{2017A&A...603A.122G} and \cite{2021MNRAS.502.2518S}). This is due to, both,  the lack of resolution (which is only partially compensated by the subgrid dynamo model in DYN5) and of the scarcity of big clusters of galaxies that can form in our $85^3 ~\rm Mpc^3$ volume.
The comparison with the approximate magnetic field constraints that can be derived from the observations listed in Sec. \ref{radio_obs} suggested first important trends:

\begin{itemize}
   \item  none of these simulations is optimal to simulate the ICM or intracluster bridges in detail, because their modest resolution prevents us from properly resolving the "MHD scale" which is key for the development of a small-scale dynamo \cite[e.g.][]{dom19}. Hence the comparison is not very informative here.

   \item in the very low-density Universe, the P and the DYN5 models have some tension with CMB limits. This problem is not as severe for astrophysical models. Vice versa, both CSF2 and CSFBH2 models produce a too low magnetisation in very low-density regions, in disagreement with blazar limits \cite[][]{blazar21}.

   \item the non-detection of diffuse emission in cluster-cluster filaments with LOFAR-HBA \cite[][]{lo21L} is consistent with all models. The non-detection of the galaxy-synchrotron cross-correlation \cite[][]{vern17} is in disagreement with the primordial model as well as on the DYN5 model because an excess cross-correlated signal is to be expected, as we discussed in \cite{gv20}.

   \item likewise, the upper limits on Faraday Rotation by \cite{2021MNRAS.503.2913A}, and even more the RM difference between pairs of giant radio galaxies by \cite{os20}, appear hard to reconcile with the large primordial seed field of the P model, as well as with the widespread amplification of magnetic fields in filaments included in the DYN5 model.

   \item the stacking detection of flat spectrum synchrotron emission between pairs of halos by \cite{vern21} is only apparently consistent with all models. However, in that work, the detailed analysis of these simulations also showed that the DYN5 and CSFBH2 models are incapable of producing enough radio emission compatible with the observed extend of the stacking excess.  In all astrophysical scenarios, some patches of high magnetic field can be found within filaments. These are typically confined to the active galaxies, from which they were released, and cannot efficiently fill the most peripheral layers of filaments, where most of the emission from the shocked gas comes from in our models.
\end{itemize}

   Based on the work mentioned above, it may be tempting to conclude that a primordial model with a seed field of $\sim 0.5 \rm ~nG$ (i.e. a primordial model similar to the P, but with initial amplitude down scaled by a factor $\sim 2$) is presently favoured by the observational data. However, there are a number of caveats:

    The statistical detections or limits to the radio signature of the cosmic web always come jointly with the complex removal of fore- and backgrounds, which are always present (and sometimes even dominant) on the large angular scales ($\geq 10'$) at which most of these observations are taken. For example, the contamination by the magnetised screen of the Milky Way is always present, and can produce artefacts \cite[e.g.][]{2020PASA...37...32H}. The removal of radio galaxies is also particularly difficult, and assessing the contribution from the background of faint or/and unresolved sources is non-trivial \cite[e.g.][]{vern21}, as well as very challenging to properly include in simulations \cite[e.g.][]{2019MNRAS.485.5285L,figaro1}. Finally, estimates coming from synchrotron emission needs a guess on the quantity of accelerated electrons to derive the amplitude of magnetic fields, but the acceleration efficiency by cosmic shocks is largely unknown \cite[e.g.][]{Bykov19,2019ApJ...876...79K,2020ApJ...897L..41X}, also based on the modelling of real radio observations of cluster radio shocks \cite[e.g.][]{2020A&A...634A..64B,rajp20,2020MNRAS.496L..48L}. Moreover, the possible additional contribution from turbulent re-acceleration operating on scales larger than cluster ``bridges" is largely unconstrained \cite[][]{bv20}.

\begin{figure}
\centering
\includegraphics[width=0.3\textwidth]{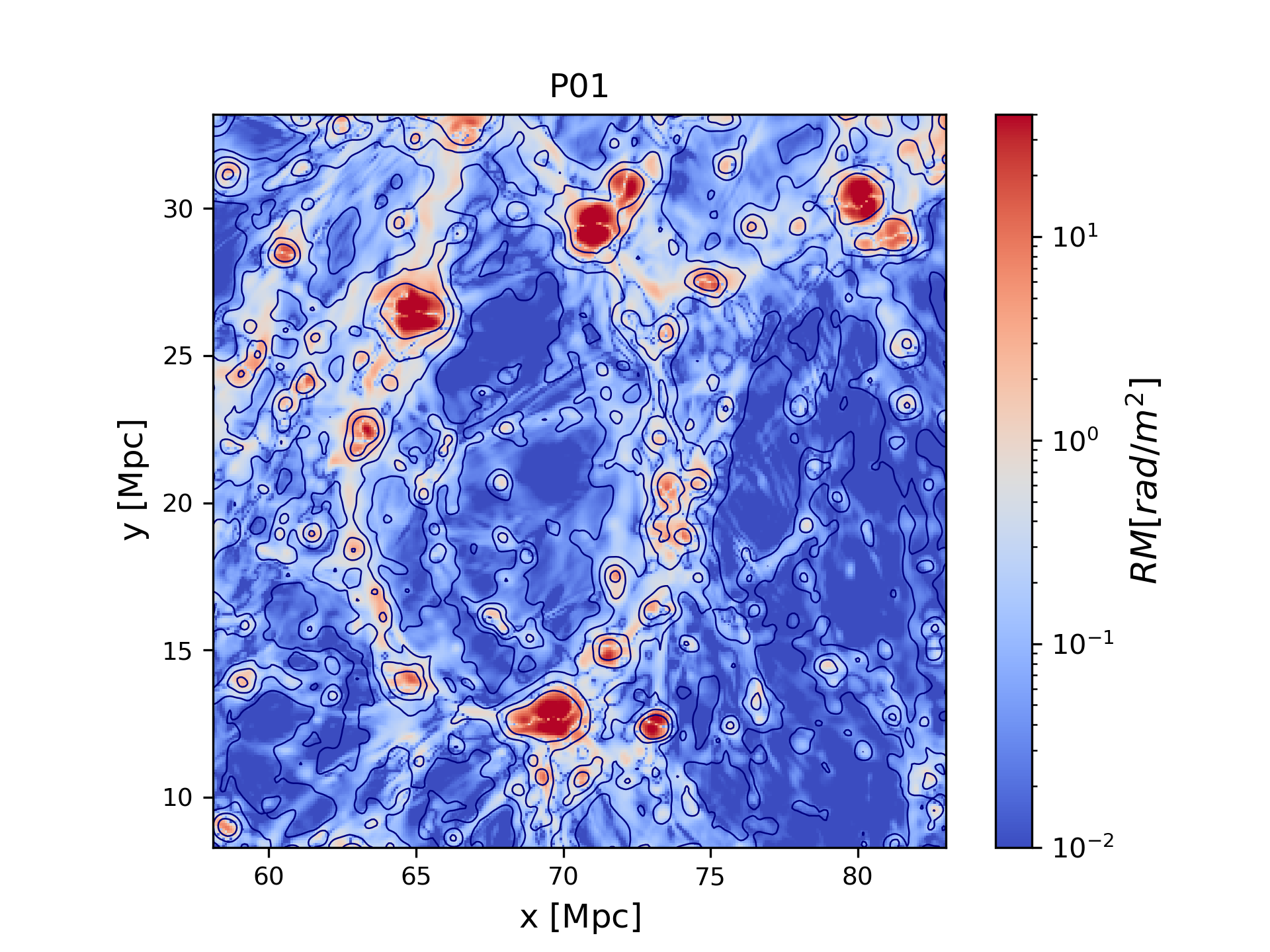}
\includegraphics[width=0.3\textwidth]{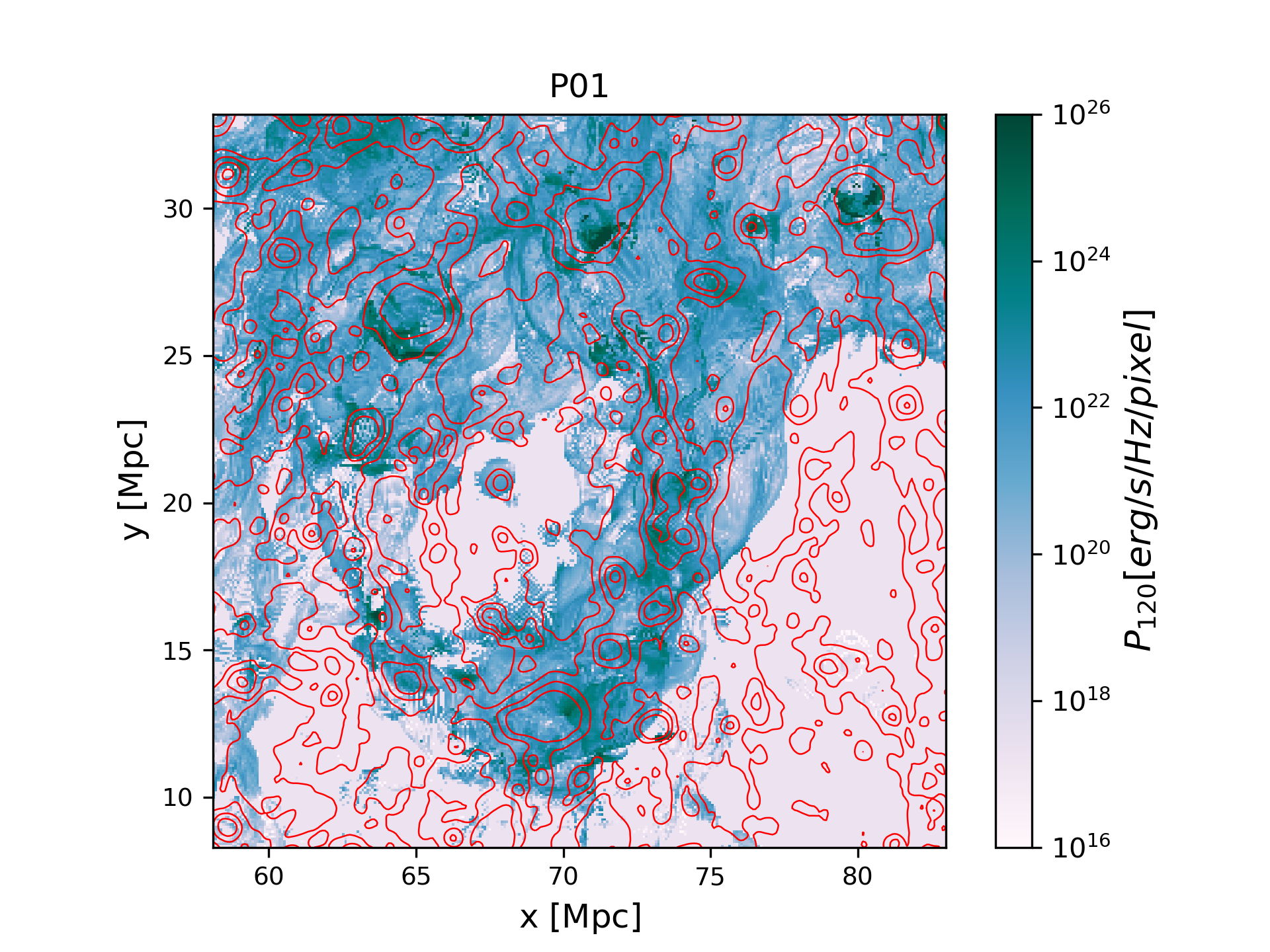}

\includegraphics[width=0.3\textwidth]{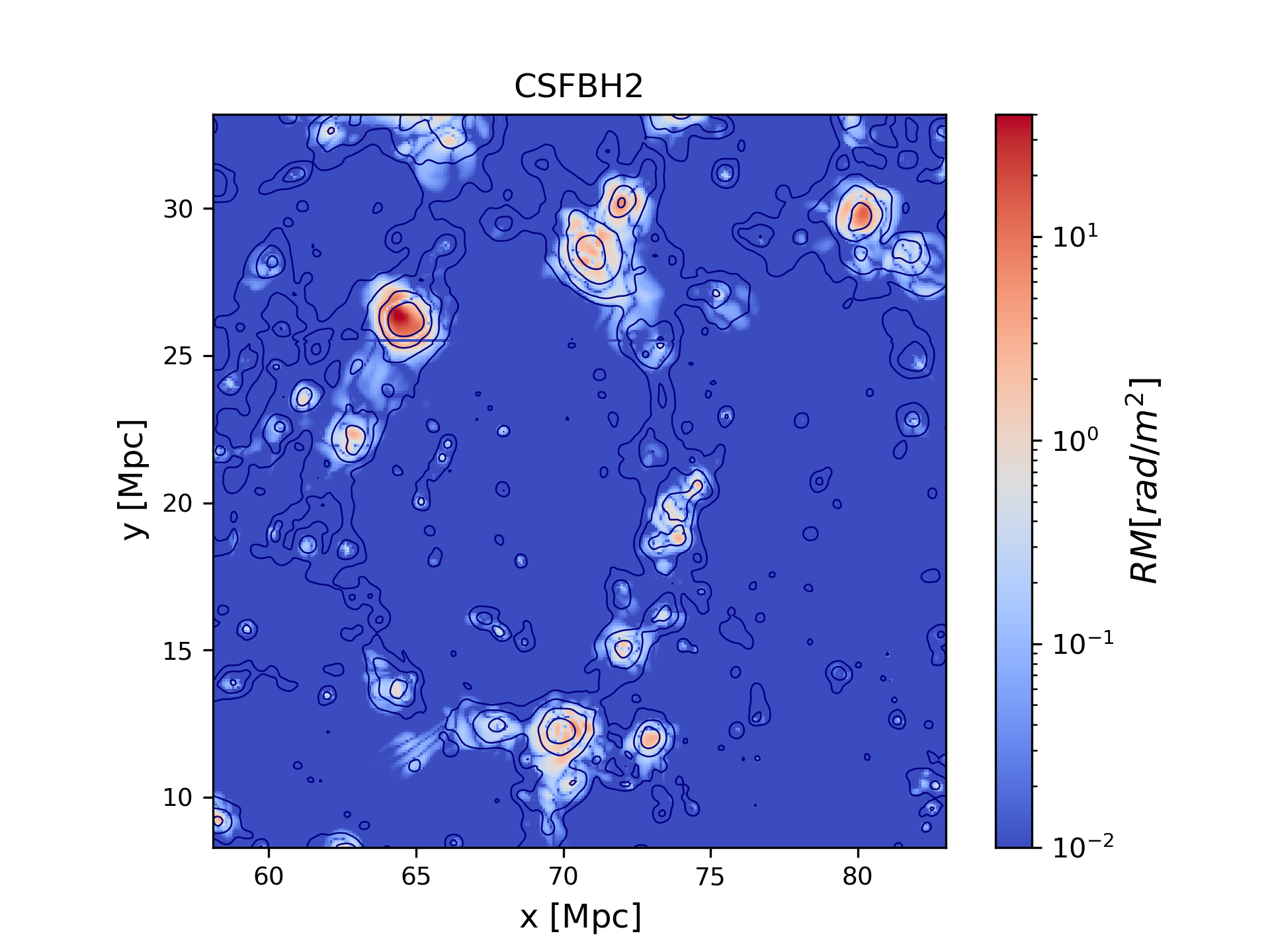}
\includegraphics[width=0.3\textwidth]{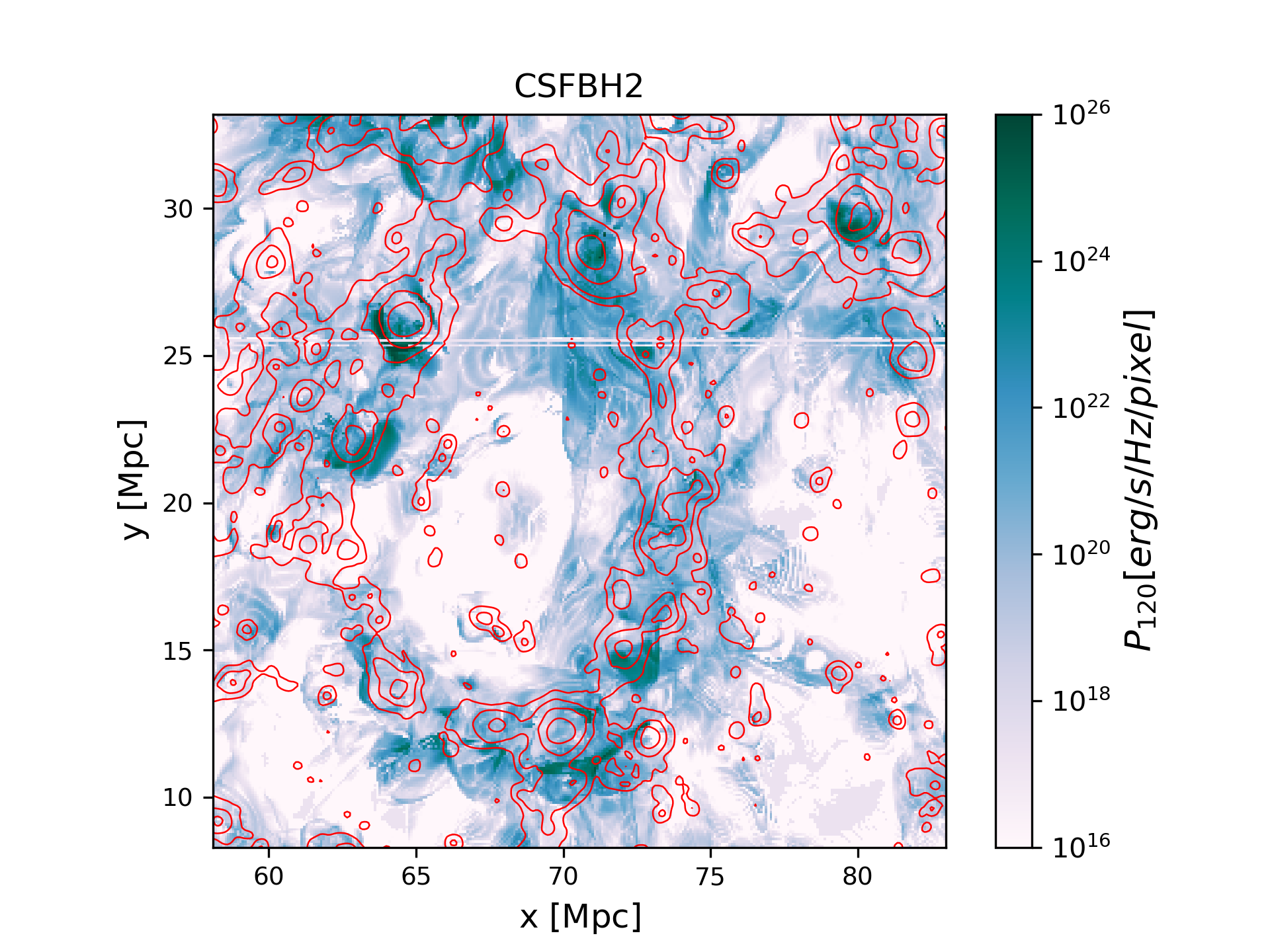}

\includegraphics[width=0.3\textwidth]{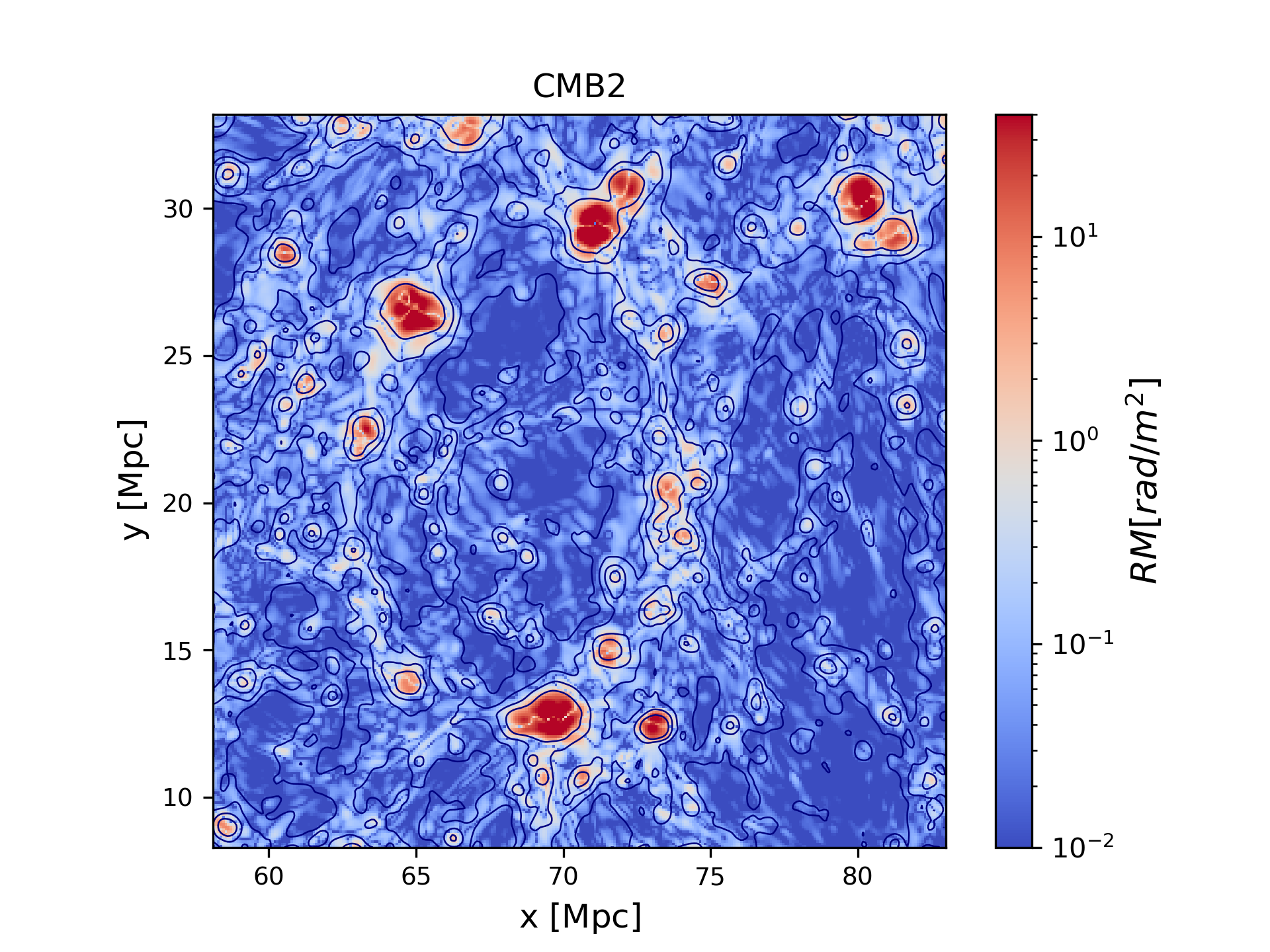}
\includegraphics[width=0.3\textwidth]{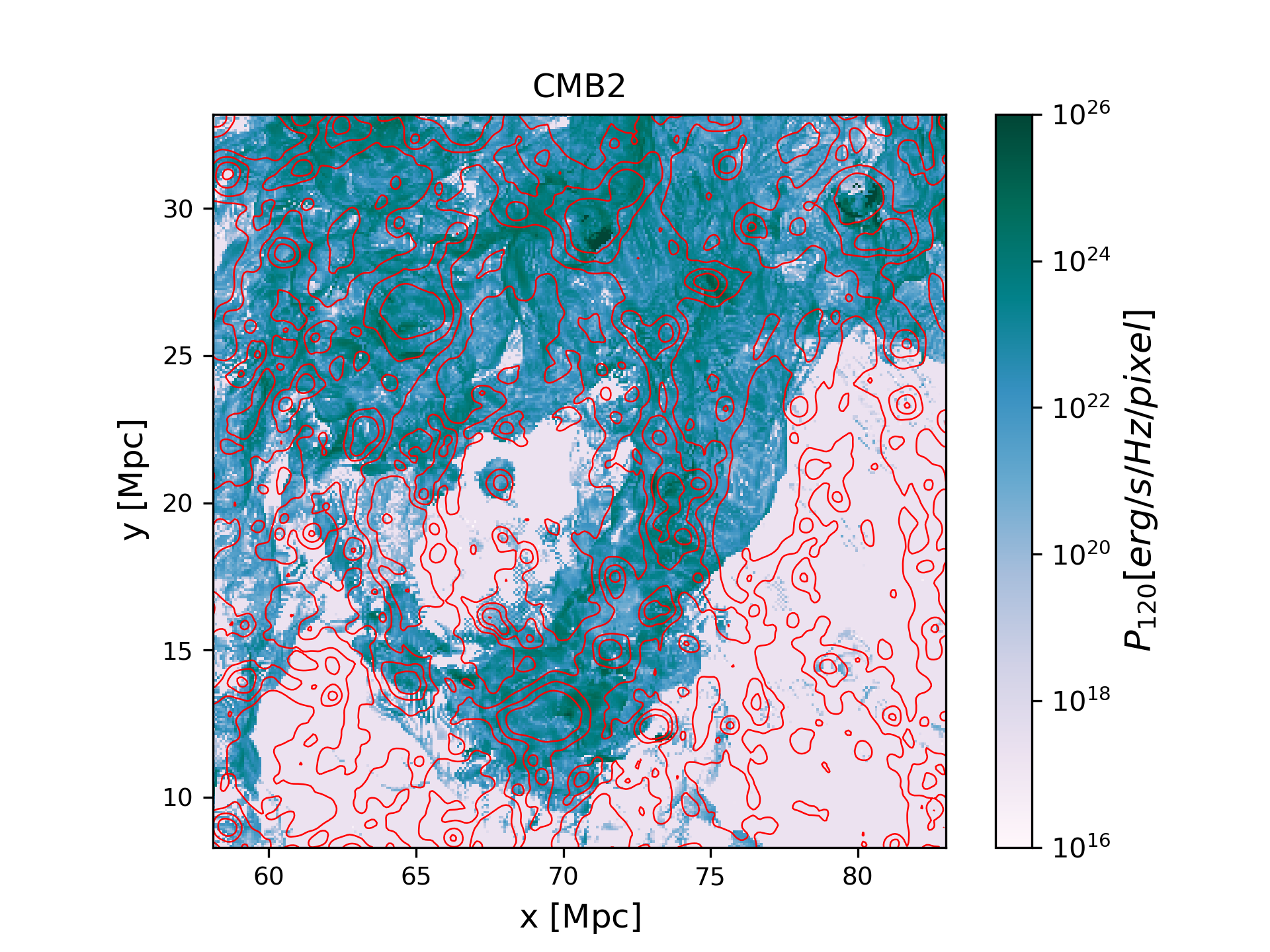}
\includegraphics[width=0.3\textwidth]{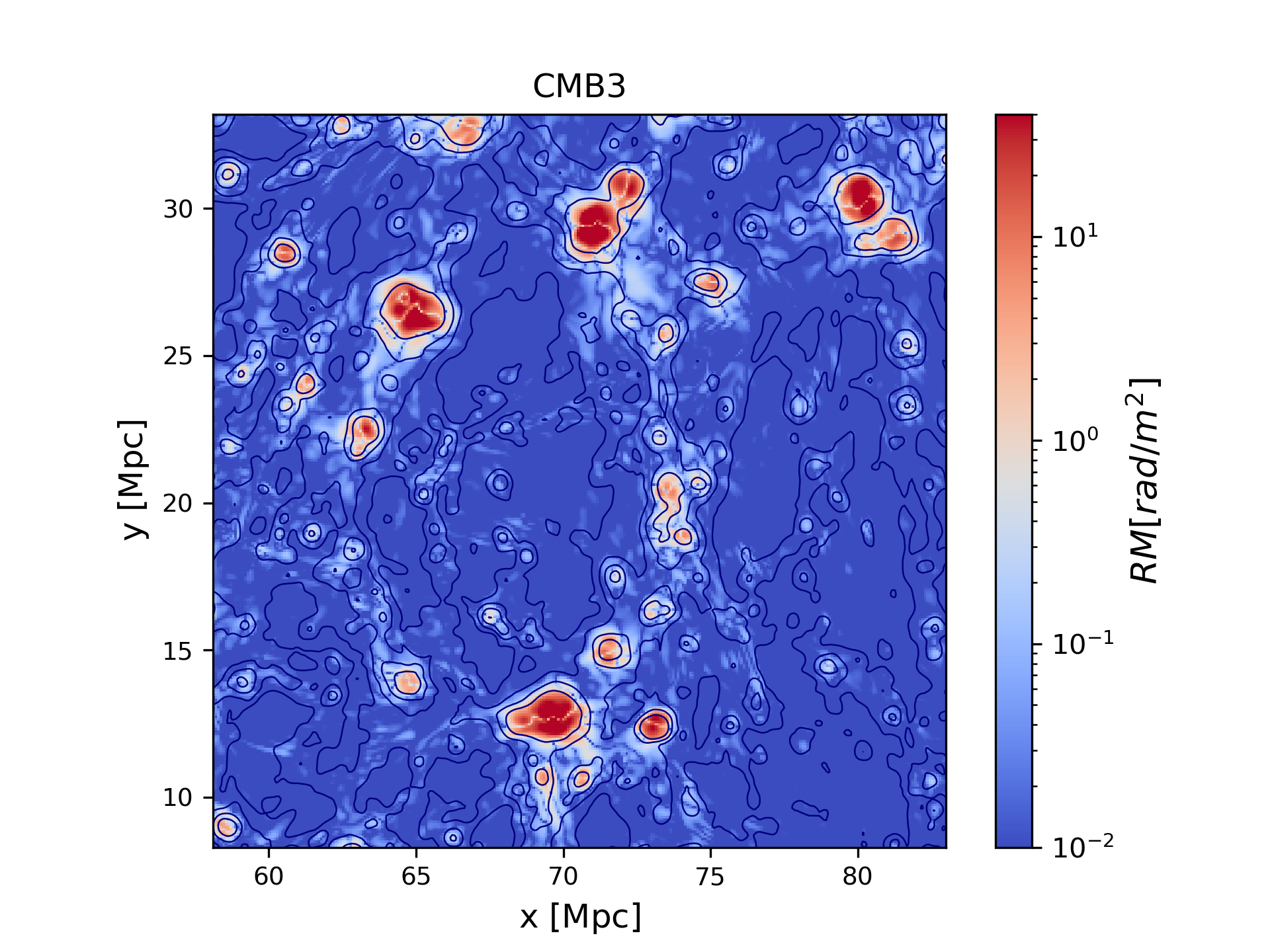}
\includegraphics[width=0.3\textwidth]{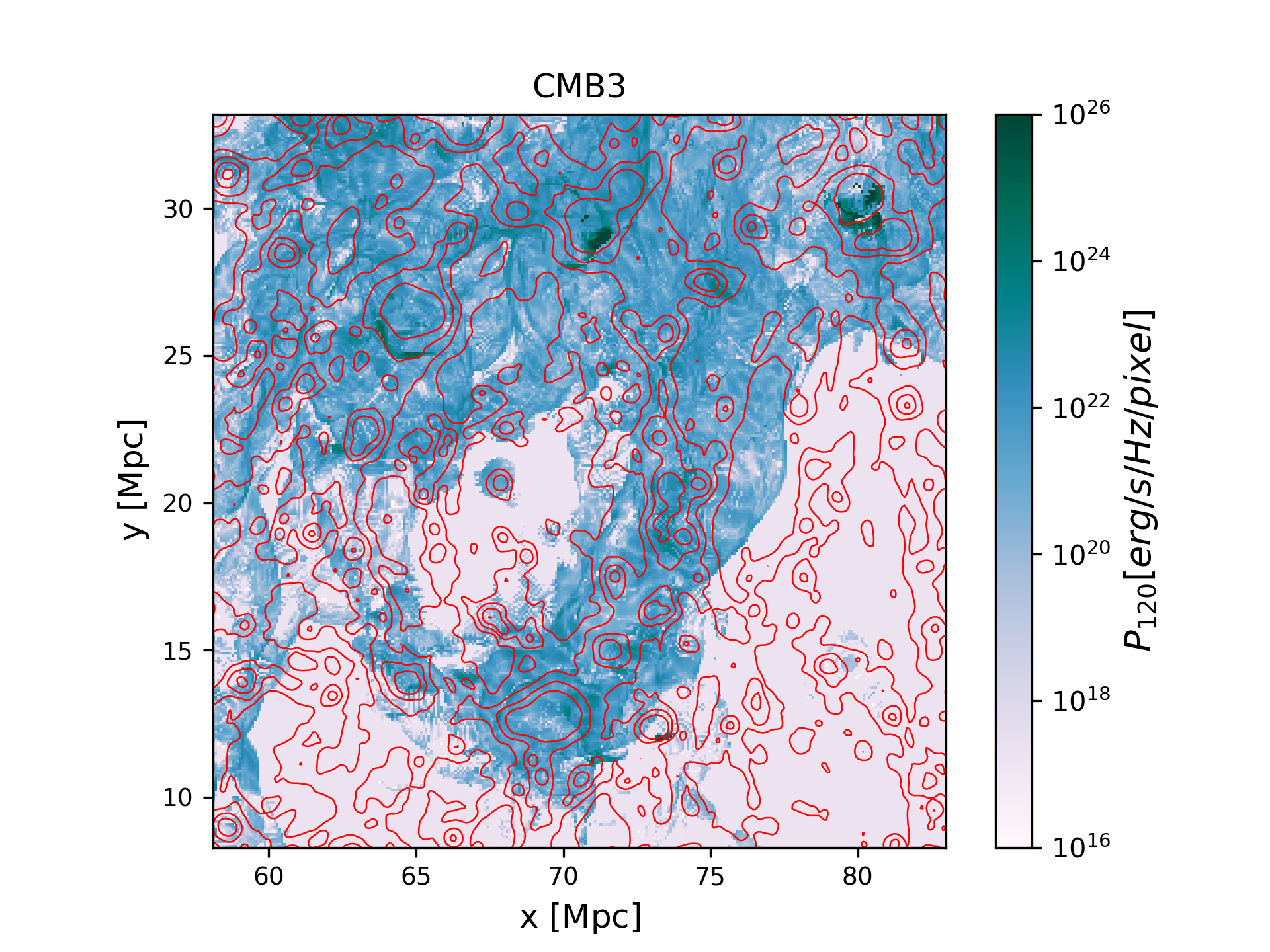}
\includegraphics[width=0.3\textwidth]{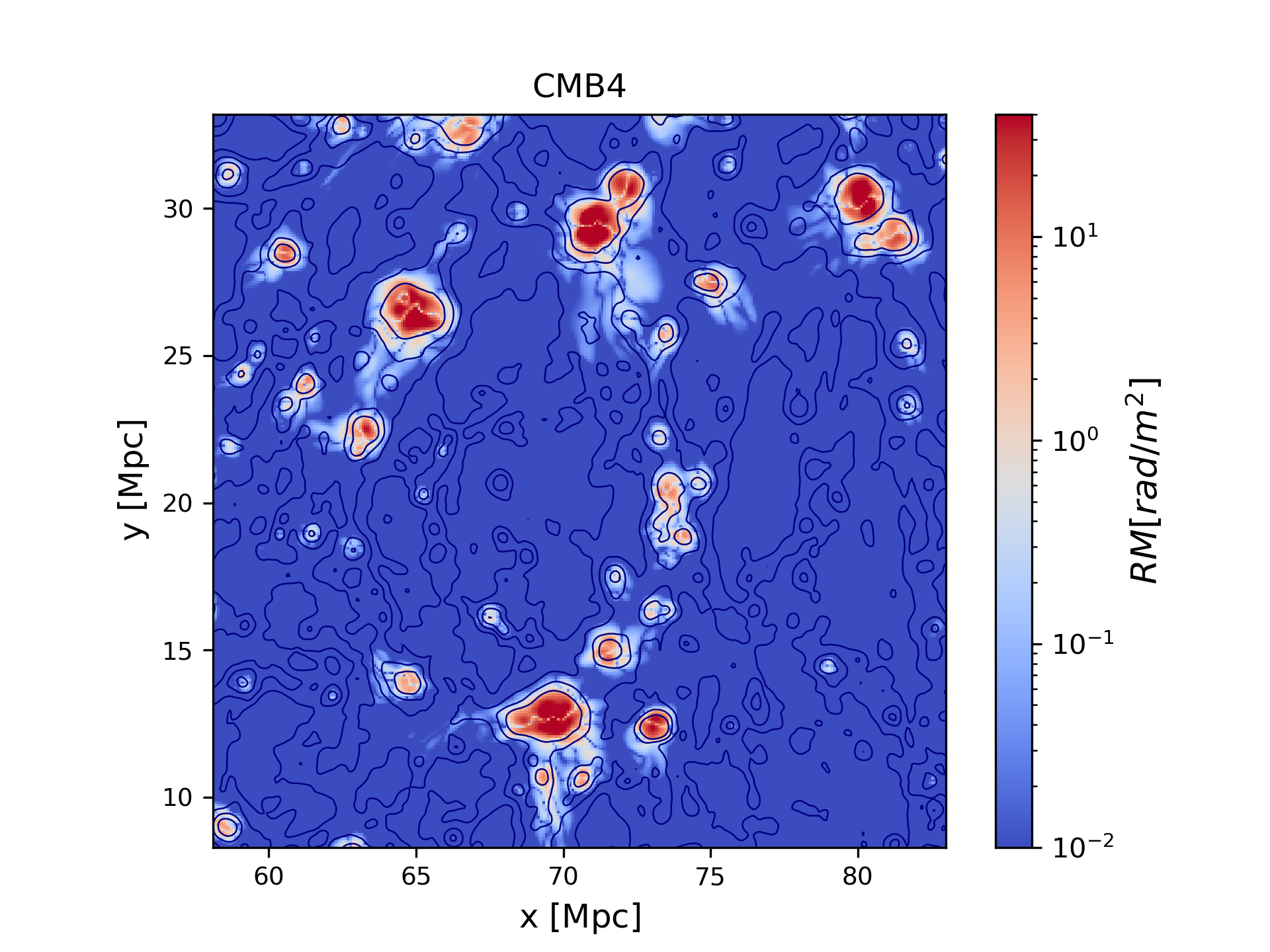}
\includegraphics[width=0.3\textwidth]{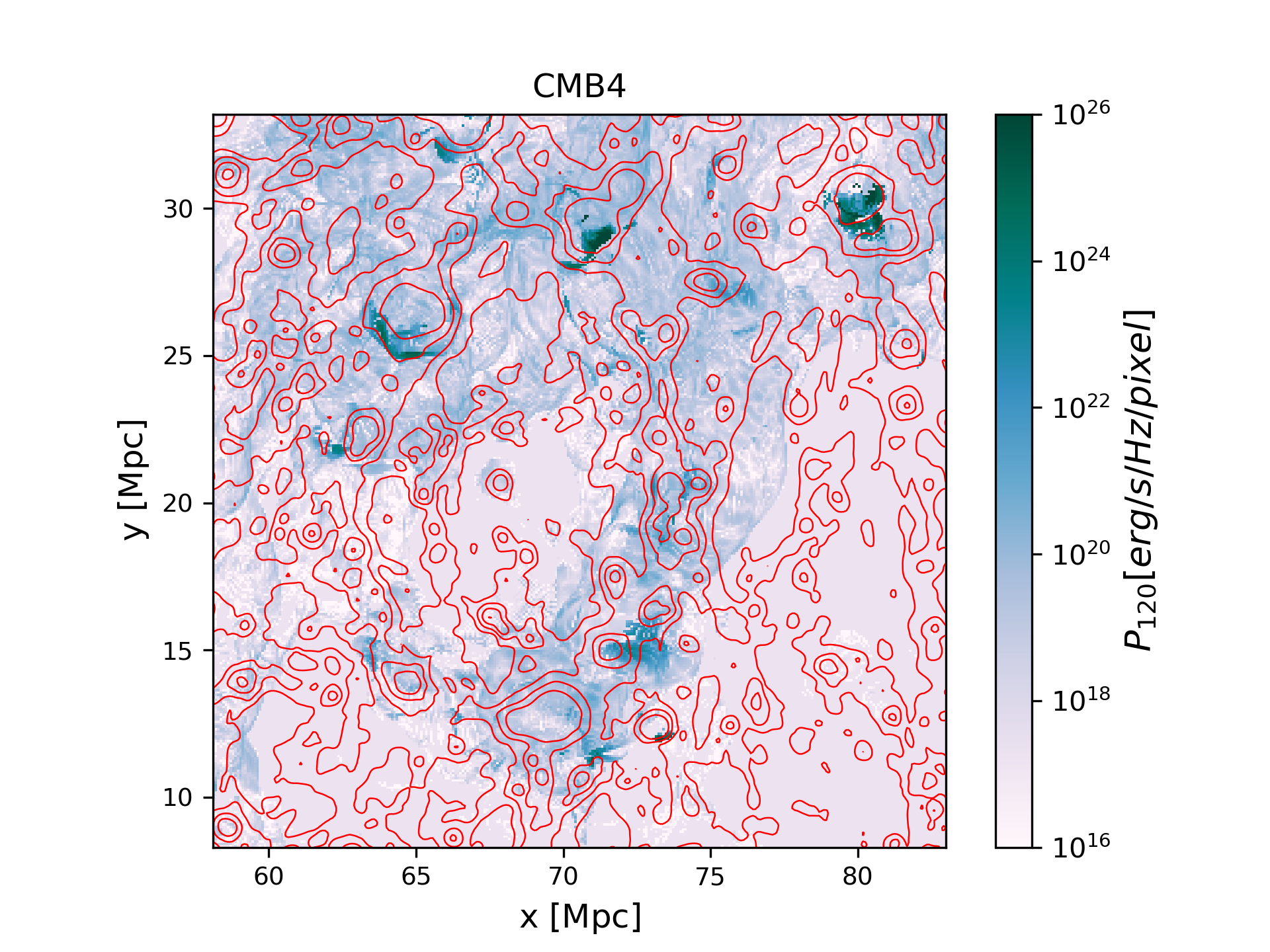}

\caption{Maps of Faraday Rotation (left) and synchrotron radio emission at 120 MHz (right), for a small portion (focusing on a few long filaments) for all our runs at $z=0.05$. The contours approximately mark the projected location of $\rho/\langle \rho \rangle \geq 10$ ordinary matter distribution.}
\label{fig:map1}

\end{figure}

   Motivated by the above comparison with observational constraints, to improve the simulated amplification of magnetic fields in halos, as well as to incorporate more realistic primordial seed fields as in  \cite{va21}, we produced our new P01, CMB2, CMB3 and CMB4 runs. Their phase diagrams are given in the last four panels of  Figure \ref{fig:phases}. Overall, these models match the boundaries of observations better than the original P model.
   % do an better job than the original P model in staying within the boundaries of observations,
   In particular, the ``blue" CMB4 model gives the most conservative (and low) average magnetic field on the scale of voids, making this model compatible at least with some of the existing limits from CMB analysis and reionization, and at the same time with limits from blazars.
   To better illustrate the different volumetric properties of these models, we give in Figure \ref{fig:map1} an example of the maps of Faraday Rotation and synchrotron radio emission at $120$ $\rm MHz$ from shock accelerated relativistic electrons (based on the same diffusive shock acceleration model used in \cite{va19}) at $z=0.05$ and integrated along a line-of-sight of $85 \rm ~Mpc$, for our most realistic primordial model (P01, CMB2, CMB3 and CMB4) and for the most complete astrophysical scenario (CSFBH2).

   The P01, CMB2 and CMB3 models show a very diffuse RM signal ($\geq 10^{-1} \rm rad/m^2$) on most scales, while the CSFBH2 and CMB4 models show signatures mostly correlated with halos. The synchrotron emission in the CSFBH2 model is patchier than in all other models, which have instead more uniform and volume filling emission patterns associated with filaments, with up to several Megaparsec from the halo they contain. It shall be noticed that the gas density structure underlying the synchrotron and RM signals (in contours) is not exactly the same in all runs.
   The presence (or absence) of feedback from galaxies and , to a lesser extent, the dynamical impact of different large-scale magnetic fields influence the gas distribution in filaments,
   as
 %  already
   tested in our
%   previous
work \cite[][]{2019MNRAS.486..981G,Banfi20}.

%\begin{figure}
%\centering
%\caption{Maps of Faraday Rotation (left) and synchrotron radio %emission at 140 MHz (right), for the same region of Fig. \ref{fig:map1}, for our runs starting from a tangled primordial magnetic field, CMB2, CMB3 and CMB4 at $z=0.05$. The contours approximately mark the projected location of $\rho/\langle \rho \rangle \geq 10$ ordinary matter distribution.}

%\label{fig:map2}

%\end{figure}

   Since several detection attempts are relying on stacking or statistical techniques, it is useful to use a sample of simulated filaments to predict the typical Faraday Rotation and synchrotron signal from the average population.

   Here we limit to low redshift data ($z \leq 0.05$, i. e. $\sim 200$ $\rm Mpc$) and  focus on the population of filaments with 3-dimensional length between $5-20 \rm ~Mpc$, which is abundant also in the small simulated volume at our disposal \cite[e.g.][]{gh15}, and which is typically targeted by direct or stacking observations of cluster-cluster pairs \cite[e.g.][]{vern21,lo21L}.

\begin{figure}
\centering
\includegraphics[width=0.3\textwidth]{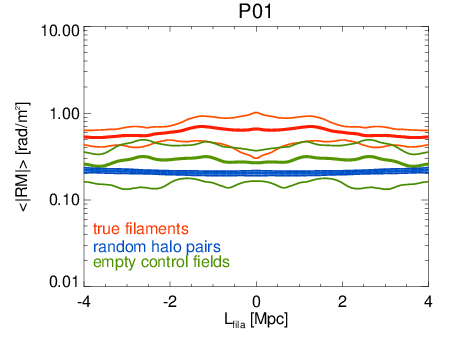}
\includegraphics[width=0.3\textwidth]{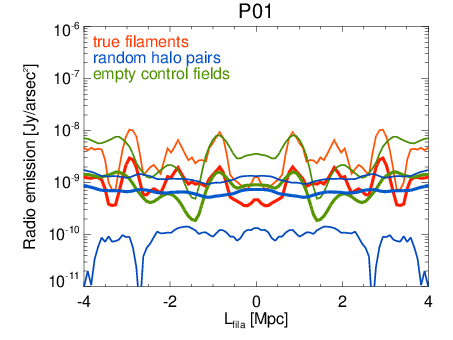}
\includegraphics[width=0.3\textwidth]{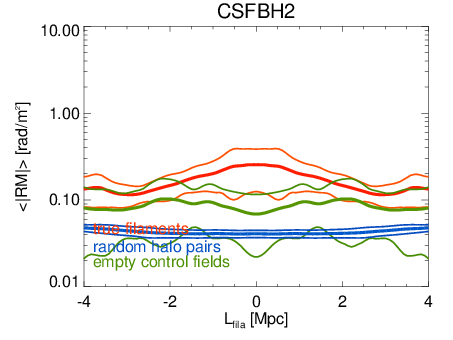}
\includegraphics[width=0.3\textwidth]{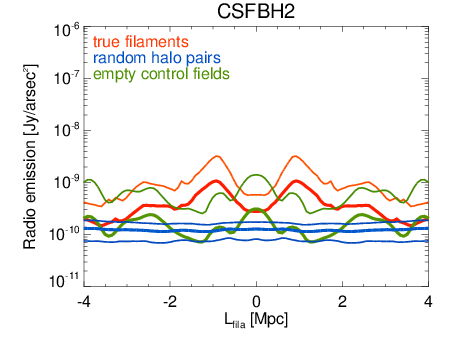}
\includegraphics[width=0.3\textwidth]{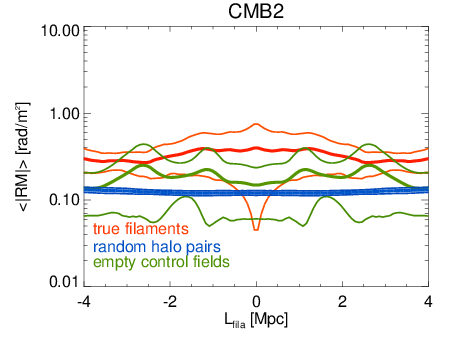}
\includegraphics[width=0.3\textwidth]{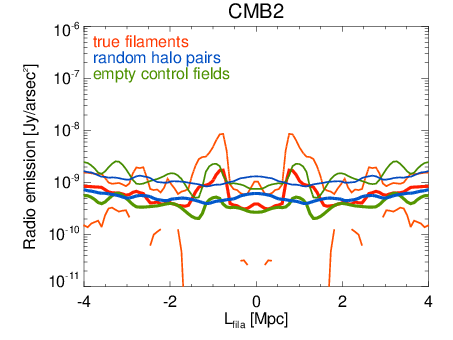}

\includegraphics[width=0.3\textwidth]{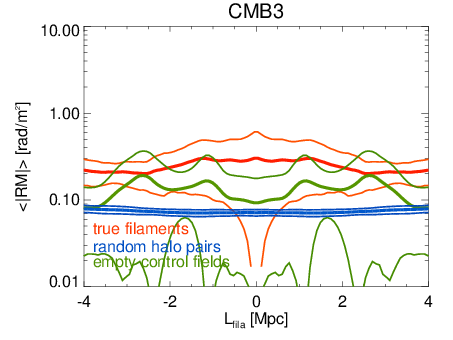}
\includegraphics[width=0.3\textwidth]{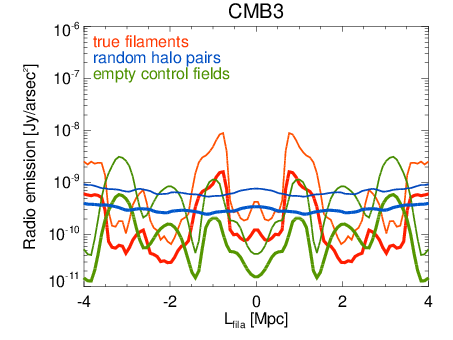}

\includegraphics[width=0.3\textwidth]{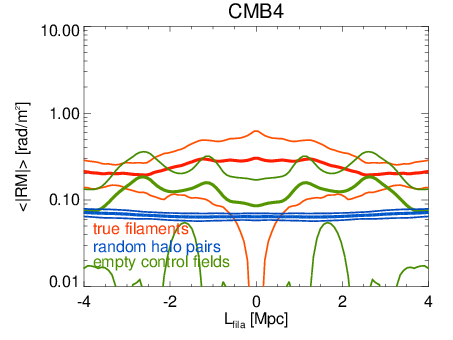}
\includegraphics[width=0.3\textwidth]{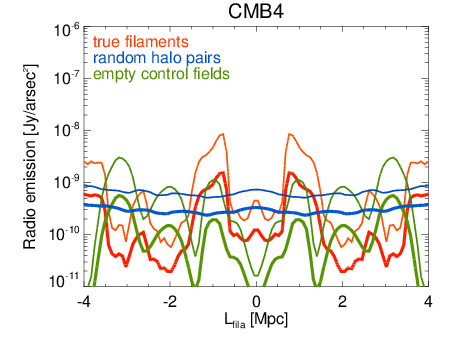}
\caption{Simulated average Faraday Rotation and radio emission from true filaments in the $5-20$ $\rm Mpc$ range of length (red), for the region between random pairs of clusters in the same range of length (blue), or for random empty control fields (green),  for all our runs. We assumed the  simulated volume at the distance of $200 \rm ~Mpc$, and we include no other foregrounds or backgrounds. Also we remove the contribution from all virial spheres of halos identified in the volume. For each model we give the sample mean and $\pm \sigma$ variance. }
\label{fig:prof1}

\end{figure}

\begin{figure}
\centering
\includegraphics[width=0.3\textwidth]{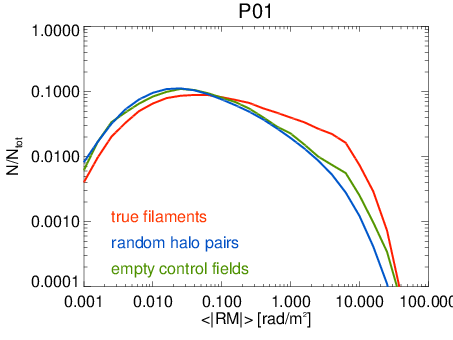}
\includegraphics[width=0.3\textwidth]{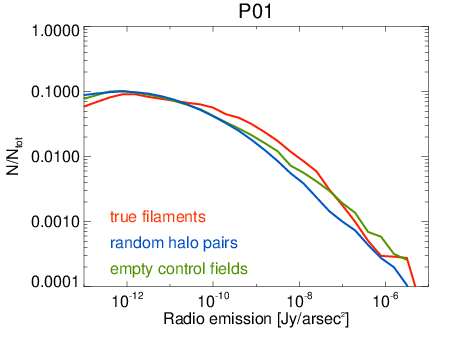}
\includegraphics[width=0.3\textwidth]{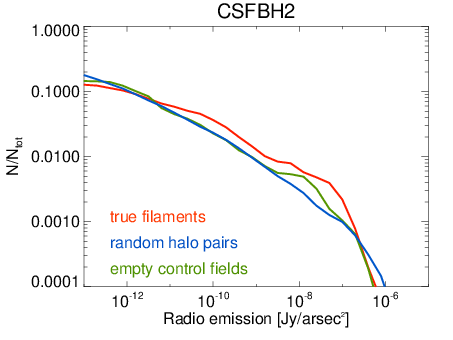}
\includegraphics[width=0.3\textwidth]{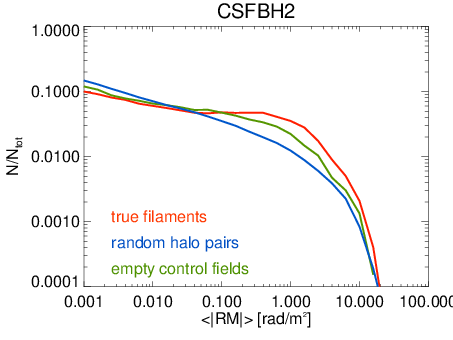}

\includegraphics[width=0.3\textwidth]{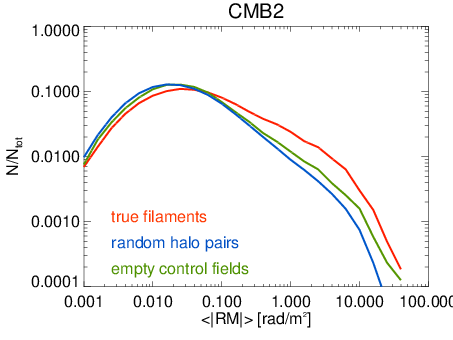}
\includegraphics[width=0.3\textwidth]{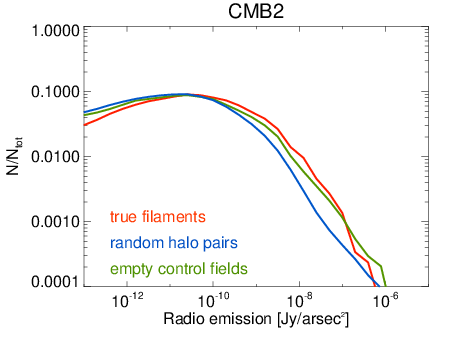}
\includegraphics[width=0.3\textwidth]{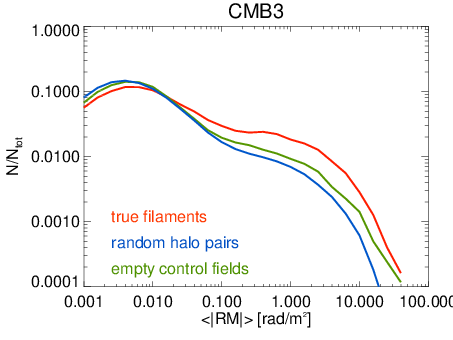}
\includegraphics[width=0.3\textwidth]{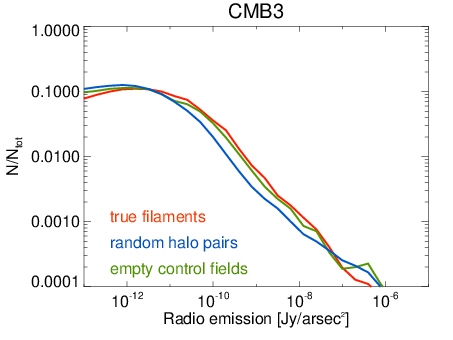}
\includegraphics[width=0.3\textwidth]{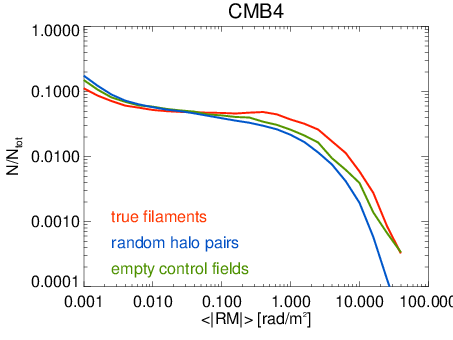}
\includegraphics[width=0.3\textwidth]{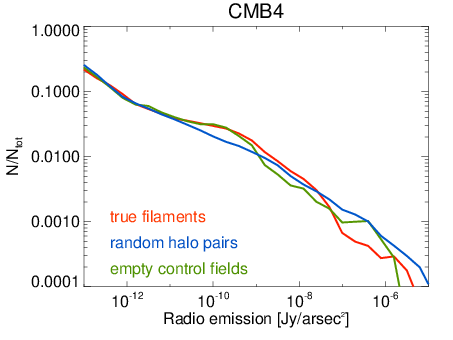}
\caption{Distribution of the  Faraday Rotation and radio emission for the ensemble of  filaments in the $5-20$ $\rm Mpc$ range of length (red), the region between random pairs of clusters in the same range of length (blue), or in random empty control fields (green), for all our runs, measured in the same data of Fig. \ref{fig:prof1}. }
\label{fig:pdf1}
\end{figure}

   First, we extract from each 3-dimensional volume a catalog of filaments in the given range of lengths, using the algorithm developed in \cite{ba21} for our simulations. By construction, all detected  filaments are delimited by $M_{\rm 100} \geq 10^{13} M_{\odot}$ halos, and are filled with gas with an average $T \geq 5 \cdot 10^5 \rm ~K$ temperature.  The RM and synchrotron emission maps are finally rescaled to the same projected angular size of $\approx 2.8^\circ$, i.e. of the order of the maximum angular scale sample by the recent stacking observation by \cite[][]{vern21}, and corresponding to a projected linear size of $10$ $\rm Mpc$ at $z=0.05$. Finally, we produced the average of both observables over the entire population of extracted filaments. For each of the $\sim 50$ filaments extracted with this procedure, we computed the average profile along a narrow stripe connecting the two nearest halos (corresponding to $\rm dx=200$ $\rm kpc$ at the reference distance) and averaged the profile with the rest of the population.
   All $\leq R_{\rm 100}$ of halos identified in the simulation are masked out from the analysis, to mimic the masking of galaxies and halos which is typically done in real observations.

   As in real observations, we also need to compare this signal with the sort of "baseline emission level" produced by the average cosmic web along the line-of-sight. We followed two approaches here: first, we generate a control set of profiles  in the same $5-20$ $\rm Mpc$ length range, by extracting pairs of halos with a {\it projected} distance in this range, which gives only $\leq 30 \%$ chances of having a true filament in between such halos \cite[][]{lo21L}.
   In a second approach, we randomly drew segments in the same length range over our projected sky model, with the contribution from halos masked out (a procedure which most often selects fields without large filaments, as it can be guessed from Figure \ref{fig:chronos}).

   Figure \ref{fig:prof1} gives the average profiles of Faraday Rotation and synchrotron emission for our simulated population of filaments, random pairs of halos or random empty fields. This attempts to quantify the realistic level of difference between magnetised filaments and control samples in a selection of our most realistic scenarios.

   In all cases, the true filaments give a higher RM and a higher synchrotron emission profile than both
   % kinds of
   control samples (halo pairs and empty fields). However, the RM signal is affected by less scatter raising the chances of detection. The middle region of filaments has a RM signal which is typically $\geq 2-3 \sigma$ above our control samples. However, the amplitude of the RM is here very low, $\sim 0.3-1 \rm rad/m^2$, which makes detection a real challenge for existing radio polarimetric observations.

   Even for targets at high galactic latitude, the
   % fluctuating
   RM fluctuations
   % contribution
   of the Galactic foreground can be as large as  $\sigma_{RM_{\rm Gal}} \sim 8 ~\rm rad/m^2$ on a  $\sim 0.5^{\circ}$ angular scale \cite[][]{2015A&A...575A.118O}. On the other hand, the typical intrinsic RM contribution from polarised background sources is $\sigma_{RM_{\rm Source}} \sim 6-12 ~\rm rad/m^2$, depending on whether the source is  a compact AGN or a star forming galaxy \cite[e.g.][]{2010MNRAS.409L..99S,banfield14}. Finally, a representative RM sensitivity that can be presently reached with the JVLA is  $\delta_{\rm RM} \approx 8 ~\rm rad/m^2$ for wide total bandwidth $\Delta\nu\simeq1$GHz observations of bright polarised sources. Future SKA-MID-like observations are expected to reach $\delta_{\rm RM}=1 ~\rm rad/m^2$ \cite[see e.g.][for a more detailed discussion]{2018Galax...6..128L}.
   On the other hand, the sample-averaged synchrotron emission presents a larger scatter, both for true filaments and for the control samples, which causes the latter to statistically overlap with the signature of true filaments within our control sample, within the scatter. Interestingly, all profiles for true filaments show a peak of emission at a distance of $\sim 3-4$ $\rm Mpc$ from their surrounding halos, which may mark the presence of substructures and shocks within such filaments, a feature which has been already noticed by \cite{figaro2}.  These features appear more pronounced when the average background magnetic field level is lower, as visible by the sequence of our CMB2-3-4 runs, and they are also related to some level of magnetic field amplification around the subhalos present in these large regions. Even if the virial volume of (sub)halos is masked out in our analysis, patches of magnetised gas can stream out of them and mix with the surrounding intergalactic medium here, increasing the scatter in the synchrotron profile.
   In any case, the average amplitude of the signal is very low, and about a  factor $\sim 10-10^2$ below what can be reached with current instruments.

   The difference between the simple uniform P01 model and the astrophysical scenario is of a factor $\sim 2$ in RM and $\sim 5-10$ in emission for such objects, while a range of values is predicted for  the different spectral models of the CMB2, CMB3 and CMB4 models. All the latter models show a large variance in the profile, the highest being associated with the ``blue" CMB4 model. This variance is caused by the large magnetic field fluctuations present in the initial conditions, which are mostly also present in the evolved filament population \cite{va21}.

   However, here we only showed the average signal of the populations, which is faint and nearly impossible to detect. Current observations can still hope to detect  $\sim 3-4 \sigma$ positive fluctuations, which can bring the previous quantities to $\sim 10 \ \rm rad/m^2$ and $\sim 1 ~\mu \rm Jy/arcsec^2$ in primordial models, e.g. at the edge of detection with present instruments, also via stacking techniques.
   This is more clearly seen in the full distribution of RM and synchrotron emission across the selected sample of filaments and control fields in the various models, shown in Fig. \ref{fig:pdf1}.
   Assessing the contribution of such large fluctuations in real stacking experiments \cite{vern21}, as well as defining strategies to build observational samples which can maximise their presence, is still a non-trivial task.

   As a final remark, it is worth stressing that the signal in {\it all} profiles shown here is a signature of the magnetic cosmic web.  , the latter represents an ubiquitous foreground and background (depending on where the main targets are located):  by definition, only gas and magnetic fields in the cosmic web contribute to the RM and synchrotron emission in our models.
   While it makes a clear detection of the radio signature of filaments only more difficult in practice \cite[e.g.][]{figaro2}, any detection of the radio signal from the matter distribution of the cosmic web (provided that other contaminants and foregrounds are removed) would be a powerful probe of cosmic magnetism \cite{lo21L}.

%%%%%%%%%%%%%%%%%%%%%%%%%%%%%%%%%%%%%%%%%%
%\section{Discussion} \label{sec:discussion}
%%%%%%%%%%%%%%%%%%%%%%%%%%%%%%%%%%%%%%%%%%
\section{Discussion \& Conclusions}
\label{sec:conclusions}

 The origin of extragalactic magnetism is still unknown. Studies of the low-density cosmic web,  outside of galaxies and clusters of galaxies, has the potential of discovering this origin because most possible scenarios for the origin of magnetic fields observed in self-gravitating halos would predict very different amplitudes and topologies of magnetic fields on the scale of filaments and voids.
Therefore, the chance of detecting an unambiguous radio signature from cosmic filaments has recently triggered many exciting observational attempts, which mostly involved low-frequency radio surveys, with LOFAR and MWA.
Given the complexity of observations (which can be variously contaminated by astrophysical backgrounds, foregrounds, outliers and contaminants) and their often statistical nature, advanced numerical simulations, such as the {\magcow} ones presented here, are often essential to turn a radio measurement into a realistic constraint on 3-dimensional magnetic fields.

At present, the non-detection of the inverse Compton cascade signal from blazars, as well as detection of the stacked signal from halo pairs seem to exclude purely astrophysical scenarios. Also primordial scenarios in which magnetic fields are maximally amplified in filaments and halos can be discarded as they would overpredict the cross-correlation between synchrotron emission and galaxies, as well as the typical RM difference between pairs of extended polarised sources. Clearly, astrophysical sources of magnetic fields related to galaxy evolution exist and are observed. However, the existing hints for large-scale signature of magnetic fields are found to require a primordial and volume-distributed seeding mechanism.

Inflationary primordial models, such as our latest simulation with tangled initial magnetic fields, are least constrained by existing observational bounds and detections.
In particular,
%we found that
%the result, that
the bluest primordial spectrum, explored here,
% that we explored
($P(k) \propto k^{1.0}$, CMB4) yields the best match with observational constraints. This match would have profound implications for the physics of the early Universe and the dynamics of inflation. More work is necessary to explore this intriguing possibility.

To conclude, it is worth stressing that a few more observable astrophysical effects are suitable to constrain the distribution of cosmic magnetic fields. We did not include them here explicitly because recent works, including our {\magcow} simulations too, have exposed many additional uncertainties related to their use.

First example, charged cosmic rays with energies $\geq 10^{18} ~\rm eV$ (aka Ultra High Energy Cosmic Rays, UHECR) are believed to have a predominantly extra-galactic origin, and their propagation towards Earth is affected by the Lorentz force  \cite[e.g.][]{Sigl:2003ay,2005JCAP...01..009D}. Large deflection angles will hamper the possibility of  locating the real sources of UHECRs at the highest energies, yet the amplitude of these deflections can greatly vary depending on model assumptions, ranging from $\leq 1^{\circ}$ to $\sim 30^{\circ}$ \cite[e.g.][]{2005JCAP...01..009D, 2014APh....53..120B,2014JCAP...11..031A}. With {\magcow} simulations, we tackled the propagation of such high energy cosmic rays,  reporting  the observed level of isotropy of $\geq 10^{18} \ \rm eV$ event can in principle be used to  rule out primordial models with a uniform magnetic field larger than $\sim 10~ \rm nG$ \cite[][]{hack16}, as well very red or very blue primordial spectra \cite[][]{va21} yet the present uncertainties in the distribution and duty cycle of sources and on the composition of cosmic rays makes such estimate still too model dependent, and degenerate with other model uncertainties \cite{hack19,2019MNRAS.484.4167G}.

Second, the RM of Fast Radio Bursts \cite[FRB, e.g.][]{2019A&ARv..27....4P}, together with their often detected Dispersion Measure, can be used to derive the average magnetic field along the line-of-sight \cite[e.g.][]{2015MNRAS.451.4277D,2017MNRAS.469.4465P,va18frb}.
However, the limited amount of FRBs with an observed RM, and the large existing uncertainties on the physical conditions at the progenitor, its location in the host galaxy, and the additional Faraday screen of our Galaxy, make it presently impossible to use FRBs as a reliable proxy of cosmic magnetism \cite[][]{hack19,hack20}, even if big progresses are expected from the SKA also here \cite[e.g.][]{2015aska.confE..92J,2015aska.confE.113T}.

Finally, also the puzzling lack of infrared absorption in the observed spectra of distant blazars \cite[e.g.][]{2012PhRvD..86h5036T,2012PhRvD..86g5024H} can be explained by the oscillation of high energy photons into axion-like particles, mediated by large scale magnetic fields \cite[][]{2008LNP...741..115M}.  We tested that a realistic distribution of $\sim 0.1- 1$ $\rm nG$ magnetic fields in voids and filaments can account for the observed spectra \cite{2017PhRvL.119j1101M}. However, such a possibility remains speculative.

To summarise, the combination of total intensity and polarimetric data at low frequencies with existing (e.g. LOFAR and MWA) and future (SKA, LOFAR2.0) large surveys
%overall
%gives us
is the best way to remove the existing uncertainties
%related to the contribution of many possible sources of signal
comin from different sources along deep cosmic lines of sight.
%, and to
Finally, low frequency observations are the best tool to  detect cosmic filaments and to measure both their magnetic content and their deep physical connection with the origin of magnetism in the Universe.

%This section is not mandatory

%%%%%%%%%%%%%%%%%%%%%%%%%%%%%%%%%%%%%%%%%%
\vspace{6pt}

%%%%%%%%%%%%%%%%%%%%%%%%%%%%%%%%%%%%%%%%%%
%% optional
%\supplementary{The following are available online at www.mdpi.com/link, Figure S1: title, Table S1: title, Video S1: title.}

%%%%%%%%%%%%%%%%%%%%%%%%%%%%%%%%%%%%%%%%%%
\acknowledgments{This study was presented by F. V.  at the "A new window on the radio emission from galaxies, clusters and cosmic web" virtual conference on 10 March 2021, during the Covid-19 pandemic.  We thank our anonymous reviewers for their helpful feedback on this manuscript. In this work we used the {\enzo} code (http://enzo-project.org), the product of a collaborative effort of scientists at many universities and national laboratories.  Our simulations were run on a) the Piz Daint supercomputer at CSCS-ETH (Lugano) b) the  JUWELS cluster at Juelich Superc omputing Centre (JSC), under projects stressicm and radgalicm; c) on the Marconi100 clusters at CINECA (Bologna), under project INA21.
We acknowledge the computing centre  of Cineca and INAF, under the coordination of the “Accordo Quadro MoU per lo svolgimento di attività congiunta di ricerca Nuove frontiere in Astrofisica: HPC e Data Exploration di nuova generazione”, for the availability of computing resources and support.
F.V., N.L., K.R., S.B., P.D.F., D.W., G.I. and M. B. acknowledge financial support from the Horizon 2020 program under the ERC Starting Grant "{\magcow}", no. 714196.
For the creation of scientific visualizations such as Fig. \ref{fig:chronos}, G.I. acknowledges the assistance of the Visit Lab of A. Guidazzoli of CINECA under the project FIBER OF THE UNIVERSE (\hyperlink{http://visitlab.cineca.it/index.php/portfolio/fiber-of-the-universe/}{http://visitlab.cineca.it/index.php/portfolio/fiber-of-the-universe/}).
S.E. acknowledges financial contribution from the contracts ASI-INAF Athena 2019-27-HH.0, ``Attivit\`a di Studio per la comunit\`a scientifica di Astrofisica delle Alte Energie e Fisica Astroparticellare'' (Accordo Attuativo ASI-INAF n. 2017-14-H.0), INAF mainstream project 1.05.01.86.10, and from the European Union’s Horizon 2020 Programme under the AHEAD2020 project (grant agreement n. 871158).
P.D.F. was partially supported by the National Research Foundation (NRF) of Korea through grants 2016R1A5A1013277 and 2020R1A2C2102800.
D.W. is funded by the Deutsche Forschungsgemeinschaft (DFG, German Research Foundation) - 441694982.
G.B. acknowledges partial support from mainstream PRIN INAF "Galaxy cluster science with LOFAR".
A.B. and C.S. acknowledge support from the ERC-StG
DRANOEL, no. 714245, and from the MIUR grant FARE SMS.  MB acknowledges support from the Deutsche Forschungsgemeinschaft under Germany's Excellence Strategy - EXC 2121 “Quantum Universe” - 390833306. DP and FF acknowledge financial support by the agreement n. 2020-9-HH.0 ASI-UniRM2 ``Partecipazione italiana alla fase A della missione LiteBIRD".
F.V. acknowledges fruitful collaboration and scientific discussion with T. Hodgson, T. Vernstrom, E. Vardoulaki,  S. O'Sullivan and V. Pomakov.}

%%%%%%%%%%%%%%%%%%%%%%%%%%%%%%%%%%%%%%%%%%
\authorcontributions{F.V. prepared the manuscript and all the numerical simulations analysed in this article. C. G. contributed to the production of numerical simulations used in this work.  G. I. produced the volume rendering used in Fig. 1. S. B. produced the filament catalog used in the Result section. D. P. produced the analytic model predictions for the CMB models simulated in this paper. All coauthors contributed in the drafting, editing and literature review of the manuscript.}

%%%%%%%%%%%%%%%%%%%%%%%%%%%%%%%%%%%%%%%%%%
%\conflictofinterests{The authors declare no conflict of interest.}

%%%%%%%%%%%%%%%%%%%%%%%%%%%%%%%%%%%%%%%%%%

\abbreviations{The following abbreviations are used in this manuscript:\\

\noindent AGN: Active Galactic Nucleus\\
AMR: Adaptive Mesh Refinement\\
CMB: Cosmic Microwave Background\\
FRB: Fast Radio Bursts\\
GLEAM: The Galactic and Extra Galactic All Sky MWA Survey\\
GPU: Graphics Processing Unit\\
HBA: High Band Antenna\\
HLL: Harten-Lax van Leer\\
ICM: Intra Cluster Medium\\
IGM: Inter Galactic Medium\\
IR: Infa Red\\
JVLA: Karl G. Jansky Very Large Array\\
KAT: Karoo Array Telescope\\
LOFAR: Low Frequency Array\\
LOS: line-of-sight\\
MAGOCW: The Magnetised Cosmic Web\\
MeerKAT: MeerKAT \\
MHD: Magneto Hydro Dynamics\\
MWA: Murchison Widefield Array\\
PLM: Piecewise Linear Method\\
RK: Runge-Kutta\\
RM: Rotation Measure\\
SDSS: Sloan Digital Sky Survey\\
SKA: Square Kilometer Array\\
SMBH: Super Massive Black Hole\\
TVD: Total Variation Dininishing\\
UHECR: Ultra High Energy Cosmic Rays\\
WHIM: Warm Hot Interagalactic Medium\\
$\Lambda$CDM: Lambda Cold Dark Matter}

\appendix
\section{Numerical simulations}

All runs presented in this paper solve
ideal  magneto-hydrodynamics on an expanding comoving volume, using a customised version of the cosmological grid code {\enzo}. The MHD scheme of preference is the  Dedner formulation \cite[][]{ded02}, i. e. a hyperbolic divergence cleaning to keep the $\nabla \cdot \vec{B}$ term as small as possible.
The MHD solver adopts the PLM (Piecewise Linear Method) reconstruction, fluxes at cell interfaces are calculated using the Harten-Lax-van Leer (HLL) approximate Riemann solver. Time integration is performed using the total variation diminishing (TVD) second-order Runge-Kutta (RK) scheme \cite[][]{1988JCoPh..77..439S}. Other work have shown that this approach is robust and competitive compared to other MHD solvers \cite[][]{kri11,2014ApJ...783L..20P,hop16}. In most runs (produced either with Piz-Daint at CSCS or with Marconi100 at CINECA) we relied on the  GPU-accelerated MHD version of {\enzo},  by \cite[][]{wang10}.

Although the use of adaptive mesh refinement is crucial to resolve the turbulent flow inside forming structures and its related dynamo amplification \cite[e.g.][]{va18mhd,2020MNRAS.494.2706Q}, for the goal of this investigations we only used a  uniform grid approach on large volumes,  as they are overall more suitable to model strong accretion shocks in low density regions, and to follow the topology of magnetic fieds in voids \cite[e.g.][]{Banfi20}.

Our main text already includes a basic description of all important adopted physical modules, but here we wish to add a few more technical details on the sub-grid dynamo model, and on our prescriptions to couple star formation and the growth of supermassive black holes to the release of magnetic fields.

\begin{itemize}
\item {\it Sub-grid dynamo amplification}:  in most runs we estimated at run-time the unresolvable amplification of magnetic fields with an approximate sub-grid approach to incorporate  small-scale dynamo amplification and overcome
the impossibility of employing adaptive mesh refinement everywhere in the simulation. This is done by measuring the gas vorticity at run-time and using it to guess the dissipation rate of solenoidal turbulence into magnetic field amplification. We assume that a small fraction, $\eta_t \approx 10^{-2}$, of such kinetic power gets channeled into the amplification of magnetic fields,  $F_{\rm turb} \simeq \eta_t \rho \epsilon_{\omega}^3/L$, where $L$ is the stencil of cells to compute the vorticity.  The fraction of turbulent kinetic power that gets converted into magnetic energy, $\epsilon_{\rm dyn}$, sets the amplified magnetic energy as $E_{\rm B,dyn} = \epsilon_{\rm dyn}(\mathcal{M})F_{\rm turb} \Delta t$.
 For a reasonable guess on $\epsilon_{\rm dyn}$, we followed the fitting formulas given by
\cite{fed14}, and set the saturation level and the typical growth time of magnetic fields as a function of the local Mach number of the flow ($\mathcal{M}$), and set $\epsilon_{\rm dyn}(\mathcal{M}) \approx (E_B/E_k) \Gamma \Delta t$, where $E_B/E_k$ is the ratio between magnetic and kinetic energy at saturation, and $\Gamma$ is the growth rate, taken from \cite{fed14}. The topology of the newly created magnetic fields is for simplicity taken to be parallel to the local gas vorticity. Manifestly,  this procedure is much simpler than more sophisticated subgrid models \cite{gr16}. However, this simplistic method reproduced the results obtained by other methods \cite[e.g.][]{ry08,va17cqg,hack19}. In these runs we adopt the same amplification efficiency calibrated in \cite[][]{va17cqg} for the DYN5 run, but in all other new runs (P, CMB2, CMB3, CMB4) we allowed the sub-grid model to be activated only for $\rho \geq 50~\langle \rho \rangle$, i. e. only within the virial radius of halos, where turbulence is predicted to be well developed and mostly solenoidal, and we increased by a factor 10 the amplification for $z \leq 1$, where the virialisation of halos is also expected to maximise the amplification of magnetic fields via small-scale dynamo \cite[e.g.][]{bm16,2020MNRAS.494.2706Q}.

\item {\it Astrophysical sources of magnetisation}: in runs including radiative (equilibrium) cooling, gas undergoes collapse and can form stars, or supermassive black holes. We resorted here to the  numerical recipes in the public version of {\enzo} \cite[][]{enzo14} to further release of magnetic dipoles, with a total magnetic energy per event which is a fixed fraction of the feedback energy.  The injection of additional magnetic energy via bipolar jets happens with an efficiency  $\epsilon_{\rm SF,b}=10\%$ with respect to the feedback energy, latter being  $\epsilon_{\rm SF} $ times the  $\dot{M} c^2$ energy accreted by star forming particles.
Likewise, we use the  prescriptions by   {\enzo}  \cite[][]{2011ApJ...738...54K} to
inject and grow SMBH and attach magnetic feedback to their thermal feedback. We assume accretion for SMBH following from the  spherical Bondi-Hoyle formula with a fixed $0.01 ~\rm M_{\odot}/yr$ accretion rate, and a fixed "boost" factor to the mass growth rate of SMBH ($\alpha_{\rm Bondi} =1000$) to balance the effect of coarse resolution, properly resolving the mass accretion rate onto our simulated SMBH particles. The injection of additional magnetic energy via bipolar jets happens with an efficiency with and $\epsilon_{\rm BH,b}=1\%$ efficiency with respect to the thermal feedback energy, which is set to be  $\epsilon_{\rm BH}$. The only run presented here which features magnetisation by stars and supermassive black hole feedback is CSFBH2, which was shown to yield the most realistic results on the cosmic star formation history, as well as on scaling relations of galaxy clusters and groups in earlier work \cite[][]{va17cqg,gv20}. We also showed results from the CSF2 run, which   only contains feedback from the star forming phase.
\end{itemize}

%In all runs we used a $\Lambda$CDM cosmological model, with density parameters $\Omega_{\rm BM} = 0.0478$, $\Omega_{\rm DM} =
%0.2602$,  $\Omega_{\Lambda} = 0.692$, and a Hubble constant $H_0 = %67.8$ km/sec/Mpc \cite[][]{2016A&A...594A..13P}.  All runs  started at $z=40$ and  have the constant spatial  resolution of  $83.3 ~\rm kpc/cell$ (comoving) and the constant mass resolution for DM of  $m_{\rm dm}=6.19 \cdot 10^{7}M_{\odot}$ per particle.

The typical  time needed to complete these runs is $\sim 100,000$ core hours for each non-radiative run and $\sim 400,000$ core hours (on average) for our runs with cooling, chemistry, star formation and SMBH particles.
%\section{}
%%%%%%%%%%%%%%%%%%%%%%%%%%%%%%%%%%%%%%%%%%

%%%%%%%%%%%%%%%%%%%%%%%%%%%%%%%%%%%%%%%%%%
%\end{paracol}
\reftitle{References}
\externalbibliography{yes}
\bibliography{franco3,dan}
\startlandscape
\end{document}